\documentclass[preprint2]{aastex6}

\usepackage{epsf}
\usepackage{graphicx}
\usepackage{natbib}
\usepackage{longtable}
\usepackage{latexsym}
\usepackage{float}
\usepackage{amsmath,esint}

\shortauthors{Suresh et al.}
\shorttitle{Automated Solar Feature Recognition and Characterization}
\begin{document}

\title{Wavelet-Based Characterization of Small-Scale Solar Emission Features at Low Radio Frequencies}

\def\IISERP{$^{1}$}
\def\Tata{$^{2}$}
\def\IISERK{$^{3}$}
\def\Haystack{$^{4}$}
\def\CalTech{$^{5}$}
\def\ASU{$^{6}$}
\def\ANU{$^{7}$}
\def\RRI{$^{8}$}
\def\Curtin{$^{9}$}
\def\MIT{$^{10}$}
\def\CfA{$^{11}$}
\def\UW{$^{12}$}
\def\Victoria{$^{13}$}
\def\UWisc{$^{14}$}
\def\UMichigan{$^{15}$}
\def\CASS{$^{16}$}
\def\CAASTRO{$^{17}$}
\def\NRAO{$^{18}$}
\def\IRINA{$^{19}$}
\def\SKA{$^{20}$}
\def\UMelbourne{$^{21}$}

\author{
A.~Suresh\IISERP,
R.~Sharma\Tata,
D.~Oberoi\Tata,
S.~B.~Das\IISERK,
V.~Pankratius\Haystack,
B.~Timar\CalTech,
C.~J.~Lonsdale\Haystack,
J.~D.~Bowman\ASU, 
F.~Briggs\ANU,
R.~J.~Cappallo\Haystack, 
B.~E.~Corey\Haystack, 
A.~A.~Deshpande\RRI, 
D.~Emrich\Curtin,
R.~Goeke\MIT,
L.~J.~Greenhill\CfA,
B.~J.~Hazelton\UW, 
M.~Johnston-Hollitt\Victoria,
D.~L.~Kaplan\UWisc, 
J.~C.~Kasper\UMichigan, 
E.~Kratzenberg\Haystack,  
M.~J.~Lynch\Curtin, 
S.~R.~McWhirter\Haystack,
D.~A.~Mitchell\CASS$^,$\CAASTRO, 
M.~F.~Morales\UW, 
E.~Morgan\MIT,  
S.~M.~Ord\Curtin$^,$\CAASTRO,
T.~Prabu\RRI, 
A.~E.~E.~Rogers\Haystack, 
A.~Roshi\NRAO, 
N.~Udaya~Shankar\RRI, 
K.~S.~Srivani\RRI, 
R.~Subrahmanyan\RRI$^,$\CAASTRO, 
S.~J.~Tingay\Curtin$^,$\CAASTRO$^,$\IRINA, 
M.~Waterson\SKA,
R.~B.~Wayth\Curtin$^,$\CAASTRO, 
R.~L.~Webster\UMelbourne$^,$\CAASTRO, 
A.~R.~Whitney\Haystack, 
A.~Williams\Curtin, 
C.~L.~Williams\MIT}
\affil{
$^{1}$Indian Institute of Science Education and Research, Pune-411008, India; \url{akshay@students.iiserpune.ac.in}\\
$^{2}$National Centre for Radio Astrophysics, Tata Institute for Fundamental Research, Pune 411007, India\\
$^{3}$Indian Institute of Science Education and Research, Kolkata-741249, India\\
$^{4}$MIT Haystack Observatory, Westford, MA 01886, USA\\
$^{5}$California Institute of Technology, Pasadena, CA 91125, USA\\
$^{6}$School of Earth and Space Exploration, Arizona State University, Tempe, AZ 85287, USA\\
$^{7}$Research School of Astronomy and Astrophysics, Australian National University, Canberra, ACT 2611, Australia\\
$^{8}$Raman Research Institute, Bangalore 560080, India\\
$^{9}$International Centre for Radio Astronomy Research, Curtin University, Bentley, WA 6102, Australia\\
$^{10}$Kavli Institute for Astrophysics and Space Research, Massachusetts Institute of Technology, Cambridge, MA 02139, USA\\
$^{11}$Harvard-Smithsonian Center for Astrophysics, Cambridge, MA 02138, USA\\
$^{12}$Department of Physics, University of Washington, Seattle, WA 98195, USA\\
$^{13}$School of Chemical \& Physical Sciences, Victoria University of Wellington, PO Box 600, Wellington 6140, New Zealand\\
$^{14}$Department of Physics, University of Wisconsin--Milwaukee, Milwaukee, WI 53201, USA\\
$^{15}$Department of Atmospheric, Oceanic and Space Sciences, University of Michigan, Ann Arbor, MI 48109, USA\\
$^{16}$CSIRO Astronomy and Space Science (CASS), PO Box 76, Epping, NSW 1710, Australia\\
$^{17}$ARC Centre of Excellence for All-sky Astrophysics (CAASTRO)\\
$^{18}$National Radio Astronomy Observatory, Charlottesville and Greenbank, USA\\
$^{19}$Instituto di Radioastronomia, Instituto Nazionale di AstroFisica, Bologna, Italy\\
$^{20}$SKA Organisation, Macclesfield SK11 9DL, UK \\
$^{21}$School of Physics, The University of Melbourne, Parkville, VIC 3010, Australia
}

\begin{abstract}
Low radio frequency solar observations using the Murchison Widefield Array have recently revealed the presence of numerous weak, short-lived and narrow-band emission features, even during moderately quiet solar conditions. These non-thermal features occur at rates of many thousands per hour in the $30.72 \text{ MHz}$ observing bandwidth, and hence, necessarily require an automated approach for their detection and characterization. Here, we employ continuous wavelet transform using a mother Ricker wavelet for feature detection from the dynamic spectrum. We establish the efficacy of this approach and present the first statistically robust characterization of the properties of these features. In particular, we examine distributions of their peak flux densities, spectral spans, temporal spans and peak frequencies. We can reliably detect features weaker than $1 \text{ SFU}$, making them, to the best of our knowledge, the weakest bursts reported in literature. The distribution of their peak flux densities follows a power law with an index of -2.23 in the $12-155$ SFU range, implying that they can provide an energetically significant contribution to coronal and chromospheric heating. These features typically last for $1-2 \text{ seconds}$ and possess bandwidths of about $4-5 \text{ MHz}$. Their occurrence rate remains fairly flat in the 140-210 MHz frequency range. At the time resolution of the data, they appear as stationary bursts, exhibiting no perceptible frequency drift. These features also appear to ride on a broadband background continuum, hinting at the likelihood of them being weak type-I bursts.  
\end{abstract}


\section{Introduction} \label{sec:intro}

The new generation radio arrays are revealing the presence of previously unappreciated variety and complexity in non-thermal solar emission features at low radio frequencies \citep{Oberoi2011, Morosan2015, TunBeltran2015}. The observations from the Murchison Widefield Array (MWA) reveal the presence of numerous short-lived, narrow-band emission features, even during what are conventionally regarded as moderate and quiet solar conditions. In terms of morphology in the MWA dynamic spectra (DS), these non-thermal features appear like miniature versions of solar type-III bursts, with spectral and temporal spans of about a few MHz and a second, respectively. Earlier radio imaging studies \citep{Oberoi2011} of such features have found their brightness temperatures to be similar to those expected for type-III bursts, implying a coherent emission mechanism behind their production. The seemingly ubiquitous presence of these features raises the possibility that they might correspond to observational signatures of nanoflares. Characterized by energies in the range of $10^{24}$--$10^{27}$ ergs, nanoflares were hypothesized by \citet{Parker1988} as a plausible solution to the coronal heating problem. At high frequencies (EUV and X-ray), the observable electromagnetic signature arises from thermal emission due to local heating of the plasma to very high temperatures by nanoflares. At low radio frequencies, the emission associated with these energetic electrons arises from coherent plasma emission mechanisms, thus, allowing even a low energy event to give rise to a much larger observational signature. This advantage makes low radio frequencies the band of choice for investigating signatures of weak coronal energy release events. In order to contribute effectively to coronal and chromospheric heating, the power law ($dN/d\mathcal{W} \propto \mathcal{W}^{\alpha}$) index, $\alpha$, of flare energies ($\mathcal{W}$) must satisfy the condition that $\alpha \leq -2$ \citep{Hudson1991}.\\

Some of the known classes of solar bursts do satisfy the $\alpha \leq -2$ requirement. \citet{Mercier1997} report an $\alpha \approx -3$ over a peak flux density range of $20-3000 \text{ SFU}$ (1 SFU = $10^4$ Jy) for type-I bursts. Type-I bursts, also referred to as radio noise storms, generally consist of short-lived ($\lesssim 1 \text{ s}$), narrow-band ($\lesssim 10 \text{ MHz}$) bursts that usually last for extended periods and are accompanied by an enhanced broadband continuum emission. Spectral and imaging observations of radio noise storms, performed by  \citet{Gergely1975} and \citet{Duncan1981}, reveal strong similarities between Type-I and decametric type-III sources. On the basis of a survey of 10,000 type-III bursts observed using the Nan{\c c}ay Radioheliograph, \citet{Saint-Hilaire2013} report a power law with $\alpha \approx -1.7$ for the distribution of peak  flux densities (in range $10^{2}-10^{4} \text{ SFU}$) of type-III bursts.  However, unlike these type-III bursts and the type-I bursts investigated by \citet{Mercier1997}, the small-scale features observed in the MWA DS are weaker with typical fluxes of about 1-100 SFU.\\

As the presence of such weak features in the MWA solar data has been established \citep{Oberoi2011} only comparatively recently, their detailed observational characteristics in terms of distributions of their spectral and temporal widths, energy content, and slopes in the frequency-time plane are yet to be determined. Such a statistical characterization of the properties of these features would be the first step towards understanding them and evaluating their contribution towards solar coronal heating. However, their high occurrence rate of thousands of features per hour in the 30.72 MHz bandwidth MWA DS necessitates an automated approach for their detection and subsequent parameter extraction from the DS. Here we present a wavelet-based automated technique for robust detection and characterization of these weak features under conditions of quiet to moderate solar activity. Though the current implementation is tuned for the MWA DS, the technique itself is more general and can be applied to DS from other instruments. As new state-of-the-art observational facilities flood the community with unprecedented large volumes of high-quality data, the need for automated data mining and analysis techniques of the sort presented here is only expected to grow more acute.\\

Section \ref{sec:data} of this paper describes the observational capabilities of the MWA and the data selected for subsequent analysis. Section \ref{sec:method} details the wavelet-based approach for automated feature detection. A statistical analysis of the properties of these features is presented in section \ref{sec:results}. The physical significance of the results obtained is discussed in section \ref{sec:discussion}. Finally, a summary of the results obtained and the conclusions from our study are presented in section \ref{sec:conclusion} of this article.

\section{Observations and Pre-Processing}\label{sec:data}

The MWA is a low frequency radio interferometer operational in the frequency range from 80-300 MHz. It is a precursor to SKA-Low and is located in the radio-quiet environment of the Murchison Radio Observatory in Western Australia. The MWA consists of 2048 dual-polarization dipoles arranged as 128 tiles, wherein each tile is a 4$\times$4 array of dipoles. For details of the technical design of the MWA, we refer readers to \citet{Lonsdale2009} and \citet{Tingay2013}. The science goals of the MWA are described in \citet{Bowman2013}. \\

The data analyzed in this work were collected using the MWA on August 31, 2014 between $\text{00:32:00 UT}$ and $\text{06:56:00 UT}$ as part of the solar observing proposal G0002. According to the SWPC event list and the NOAA/USAF Active Region Summary  (\url{http://www.solarmonitor.org}) for this day, this observing period was marked by medium levels of solar activity with occurrence of one B-class flare (B8.9 at $\text{03:51:00 UT}$) and two C-class flares (C1.3 at $\text{01:51:00 UT}$ and C3.4 at $\text{05:37:00 UT}$, both from the active region with NOAA number 12149). A type-III solar radio burst was also reported to occur at $\text{01:25:00 UT}$ on this day.\\

The data were taken in a loop cycling from $79.36 \text{ MHz}$ to $232.96 \text{ MHz}$ in 5 steps of $30.72 \text{ MHz}$, spending 4 minutes at each frequency band. The entire 30.72 MHz bandwidth in each data set is comprised of 24 coarse spectral channels, each $1.28 \text{ MHz}$ wide. Each coarse spectral channel is further composed of 32 fine spectral channels with a resolution of 40 kHz each. The time resolution of the data collected is 0.5 seconds.  The MWA interferometric data above $100 \text{ MHz}$ is flux-calibrated according to the prescription developed by \citet{Oberoi2016} and \citet{Sharma2017}. This flux calibration technique provides estimates of the solar flux densities and brightness temperatures by accounting for known contributions from the sky, the receiver, and ground pickup noise to the system temperature. The receiver temperatures and ground pickup temperatures are obtained from a mix of laboratory and field measurements. Estimates of the sky temperature are obtained using the \citet{Haslam1982} 408 MHz all-sky map, scaled with a spectral index of 2.55 \citep{Guzman2011}, as a sky model. The need to keep the Sun unresolved for application of this flux calibration technique constrains us to using only short baselines. This non-imaging study uses data from one such short baseline of physical length 23.7 m between tiles labeled ``Tile011MWA'' and ``Tile021MWA''. The outputs from the flux calibration technique described in \citet{Oberoi2016} form the inputs for our study. Here, we present the analysis for data collected in the XX polarization alone, that for the YY polarization is analogous.   

\section{Methodology}\label{sec:method}
\begin{figure*}
\figurenum{1}
\epsscale{1.16}
\plottwo{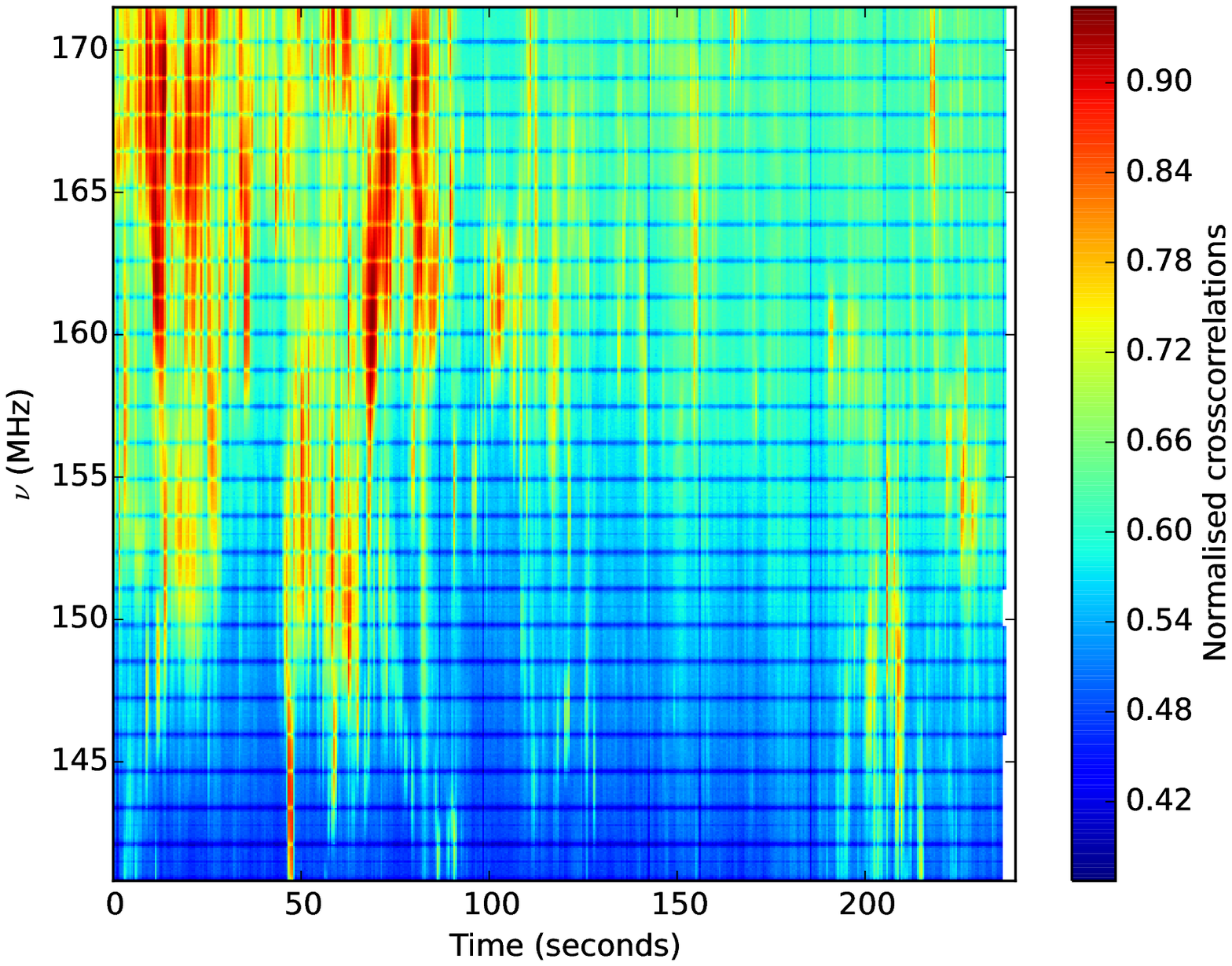}{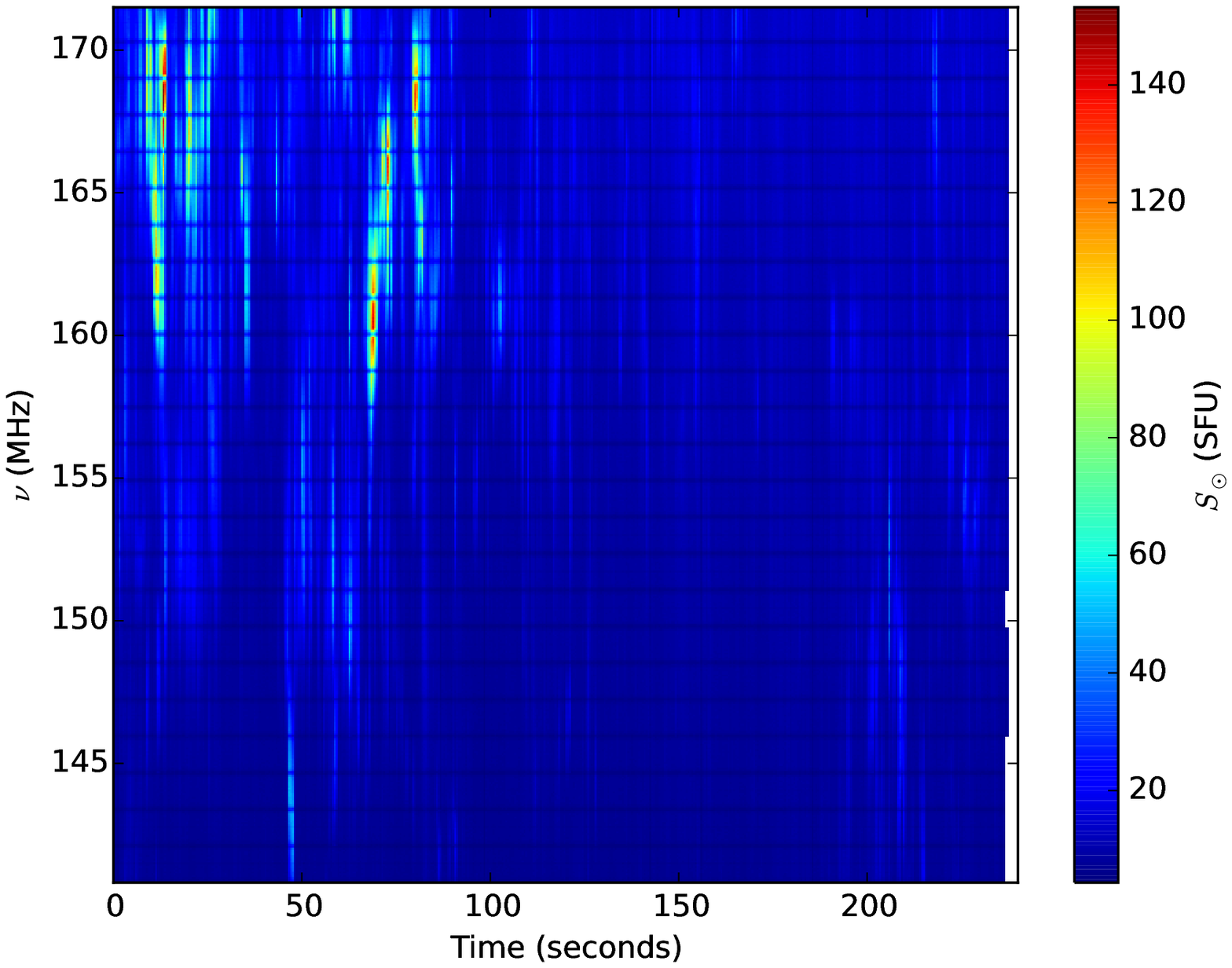}
\caption{Left: A sample MWA DS of normalized cross-correlations. Right: Flux-calibrated version of the same DS.\label{fig:f1}}
\end{figure*}

Figure \ref{fig:f1} depicts a sample raw MWA DS of normalized cross-correlations on the left and its flux-calibration version on the right. The features of interest in this work appear as short-lived, narrow-band vertical streaks against a broadband background continuum. 

\subsection{Removal of instrumental artifacts}\label{instrumental_artifact_removal}

The horizontal features are instrumental artifacts arising due to the poor instrumental response at the edges of coarse spectral channels and need to be removed. These artifacts are corrected for by performing linear interpolation across the systematics-affected channels. As the coarse channel edges at the very start and end of the observing band cannot be corrected by interpolation, these are simply discarded. Recording glitches sometimes affect the beginning and end of data recording. To avoid contamination from such issues, we routinely discard the first six and the last nine time slices of data as well.\\

\begin{figure*}
\figurenum{2}
\epsscale{1.16}
\plottwo{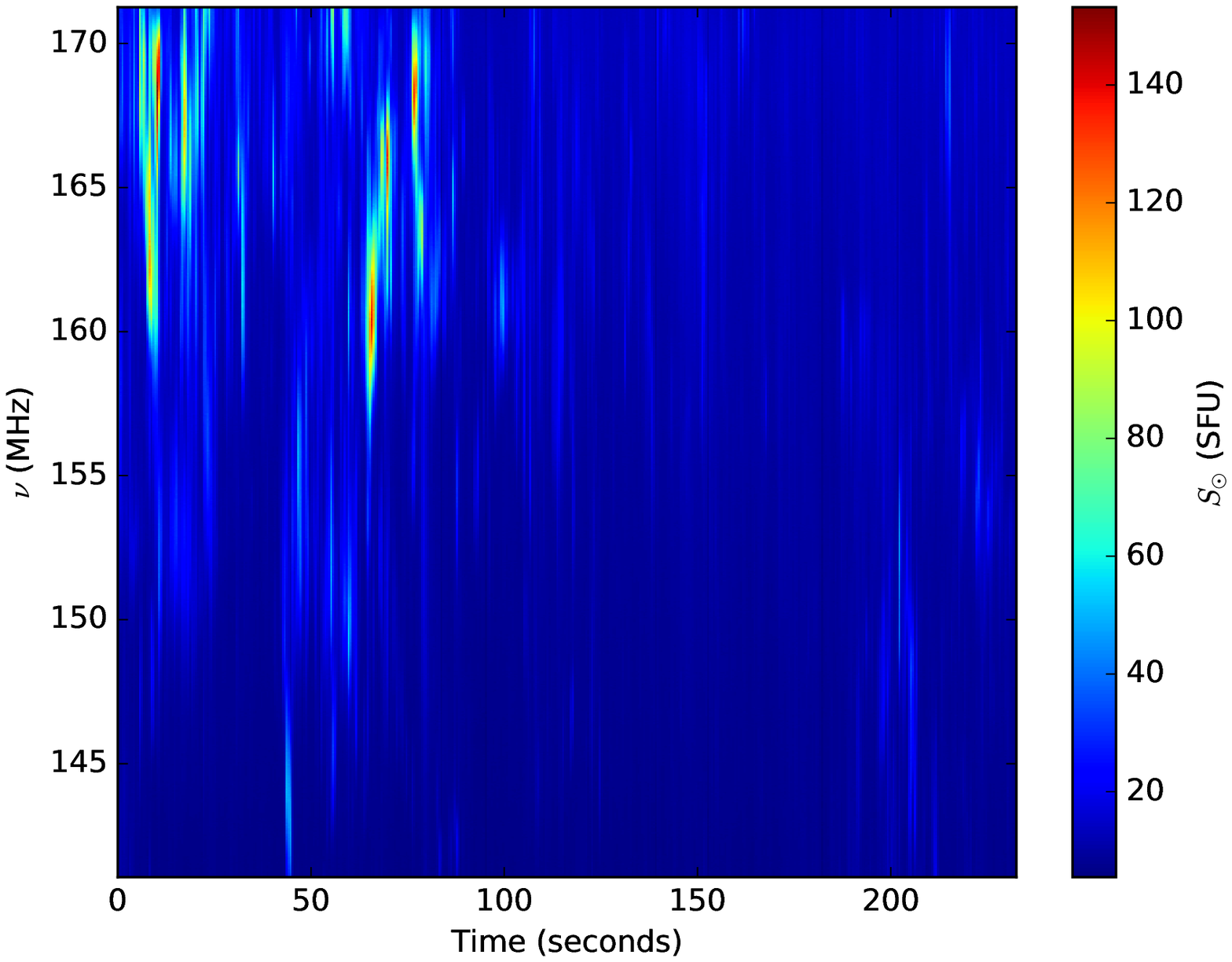}{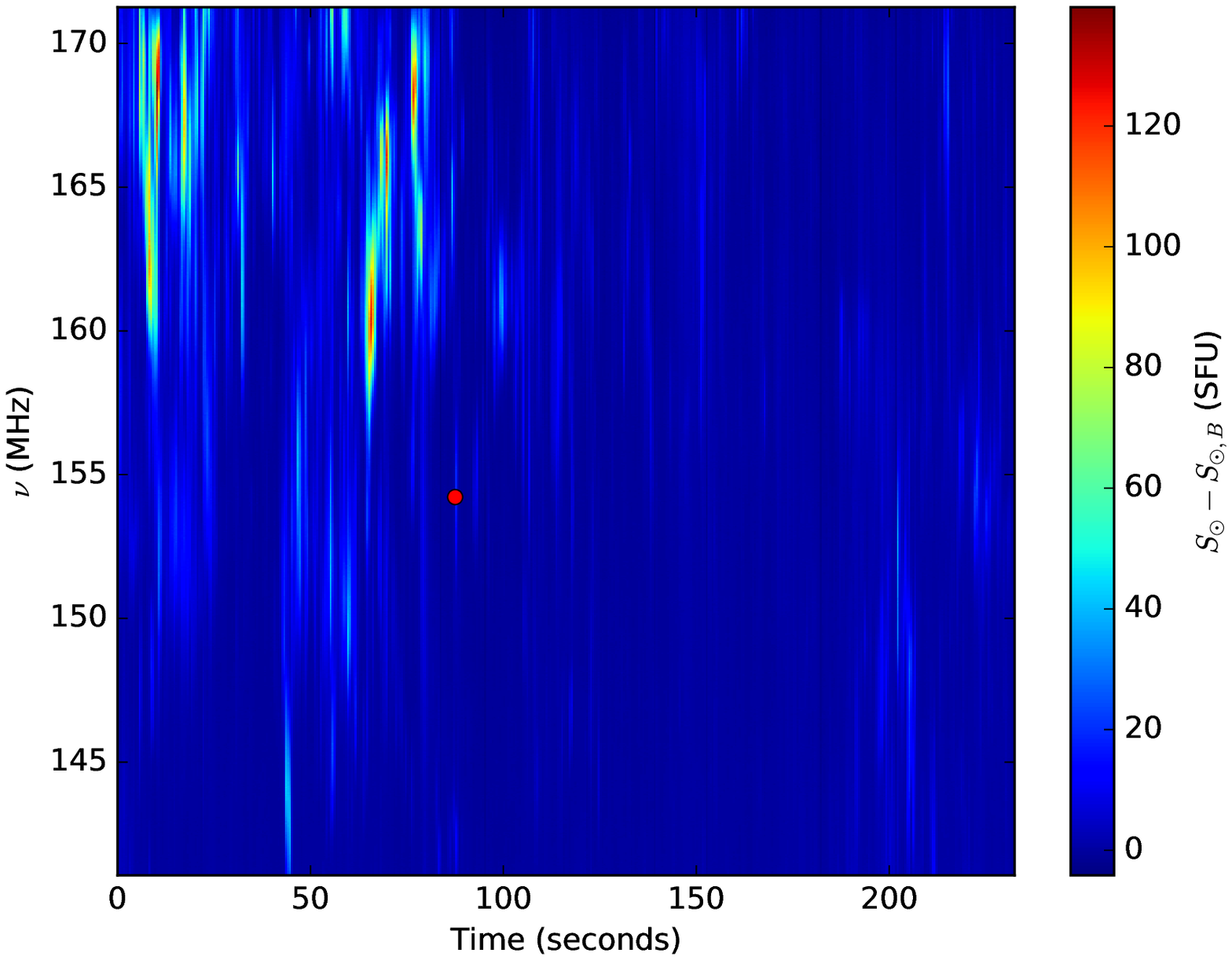}
\caption{Left: A version of the flux calibrated DS free of instrumental artifacts. Right: Background-subtracted version of the same DS. Features of interest can be easily identified in this processed DS. We note that these features also overlap in many instances. One feature that appears to be relatively isolated from the others is marked by a red circle in the right panel.\label{fig:f2}}
\end{figure*}

Though the MWA is located in a region with very little radio frequency interference (RFI), radio waves reflected from aircraft can occasionally interfere with the radio signals picked up by a tile and thereby, corrupt the data collected. Manual RFI-flagging followed by linear interpolation across RFI-affected segments of the DS is carried out to ensure a RFI-free DS for efficient feature detection. The left panel of Fig. \ref{fig:f2} displays an instrumental artifact-free version of the DS shown in Fig. \ref{fig:f1}. 

\subsection{Background continuum subtraction}\label{background_removal}
The solar radiation can be thought of as a superposition of  sporadic non-thermal radio features with a spectrally varying, broadband background continuum. Spectral variations in the background flux density can often distort the spectral profiles of features in the DS. For improving our efficiency at picking up small-scale features from the DS, it is, therefore, necessary to disentangle spectral flux density variations arising from these features from that associated with the background. \\
\begin{figure*}
\figurenum{3}
\gridline{\fig{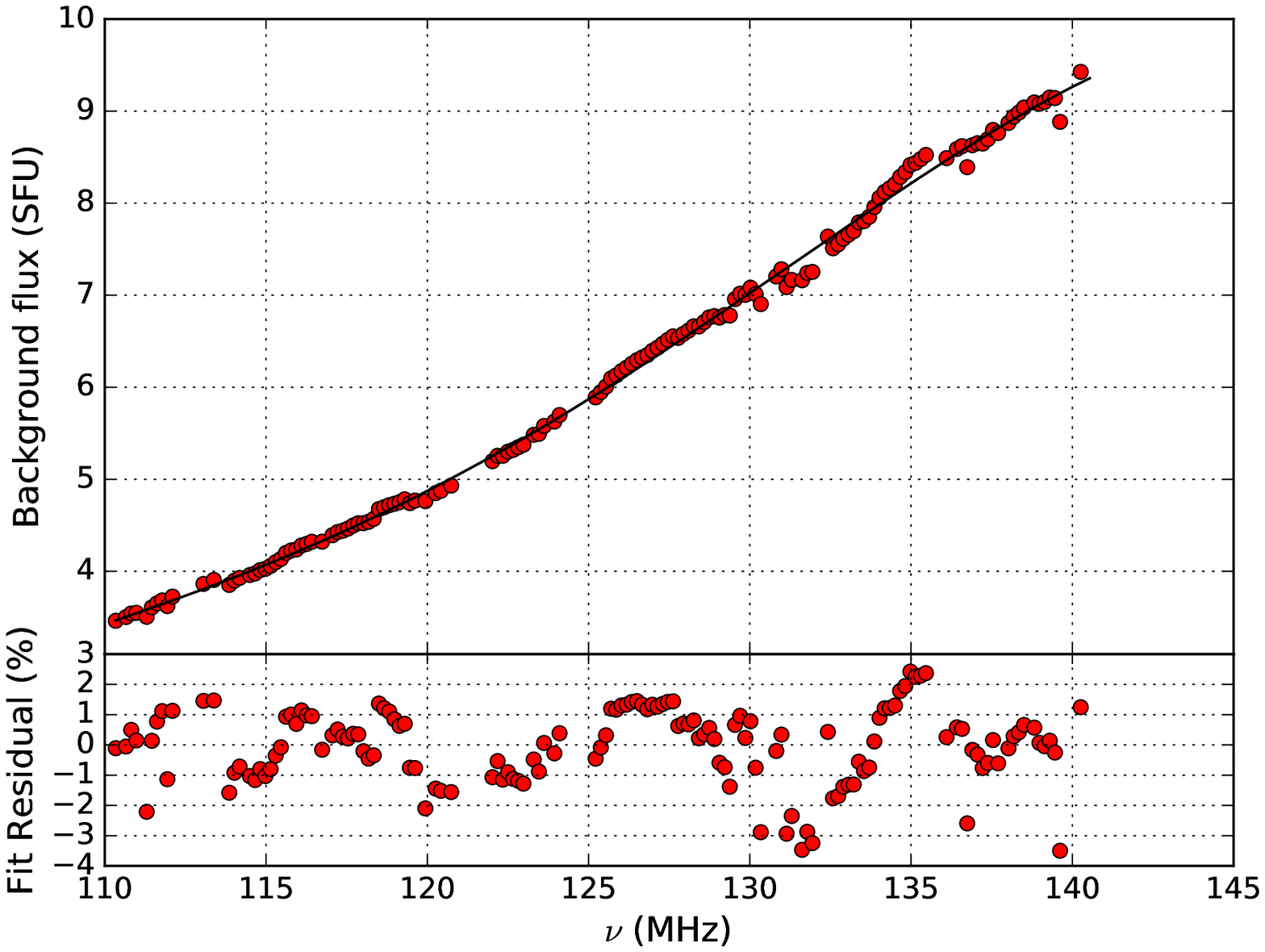}{0.5\textwidth}{(a)}
          \fig{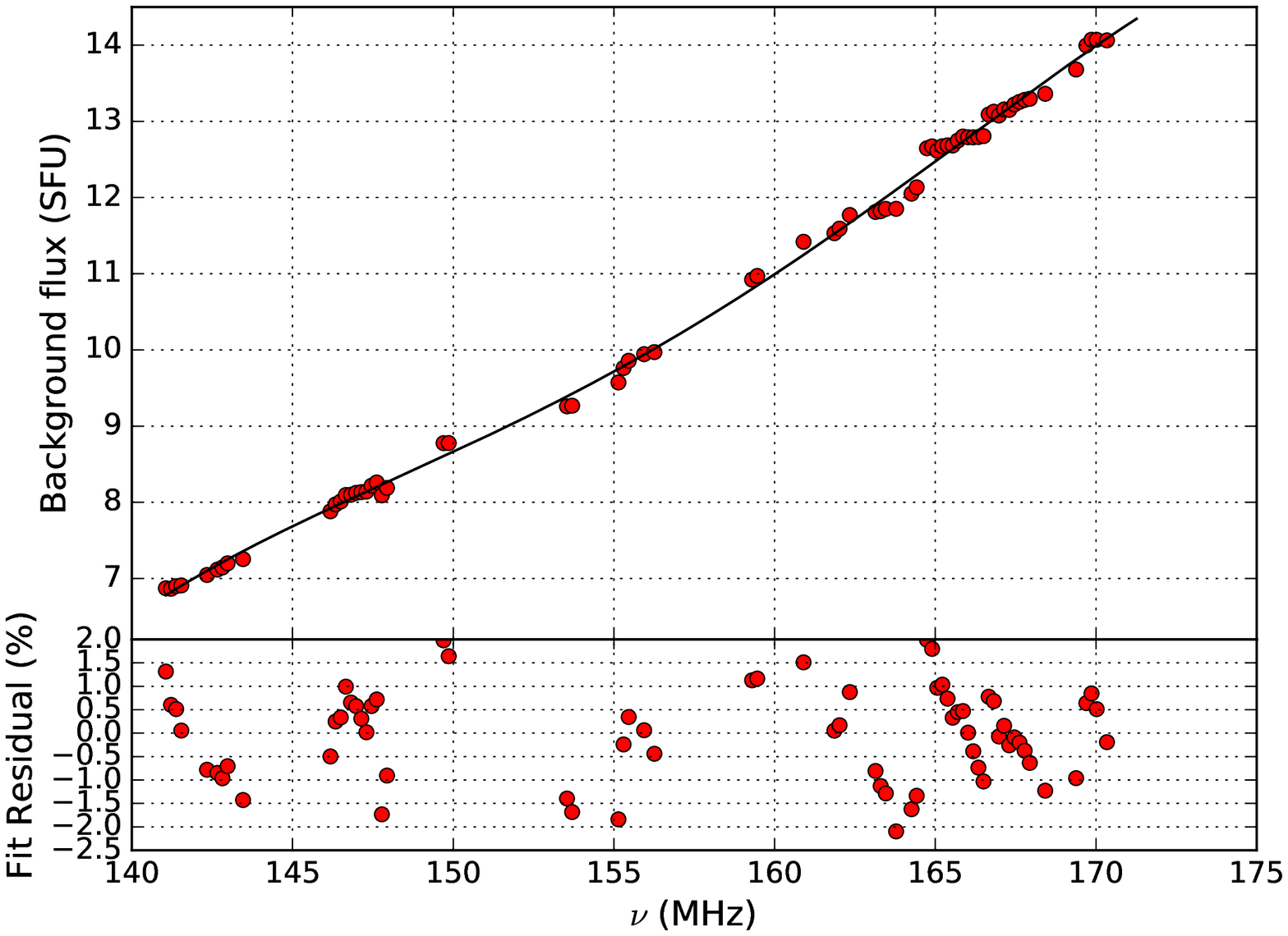}{0.5\textwidth}{(b)}
          }
\gridline{\fig{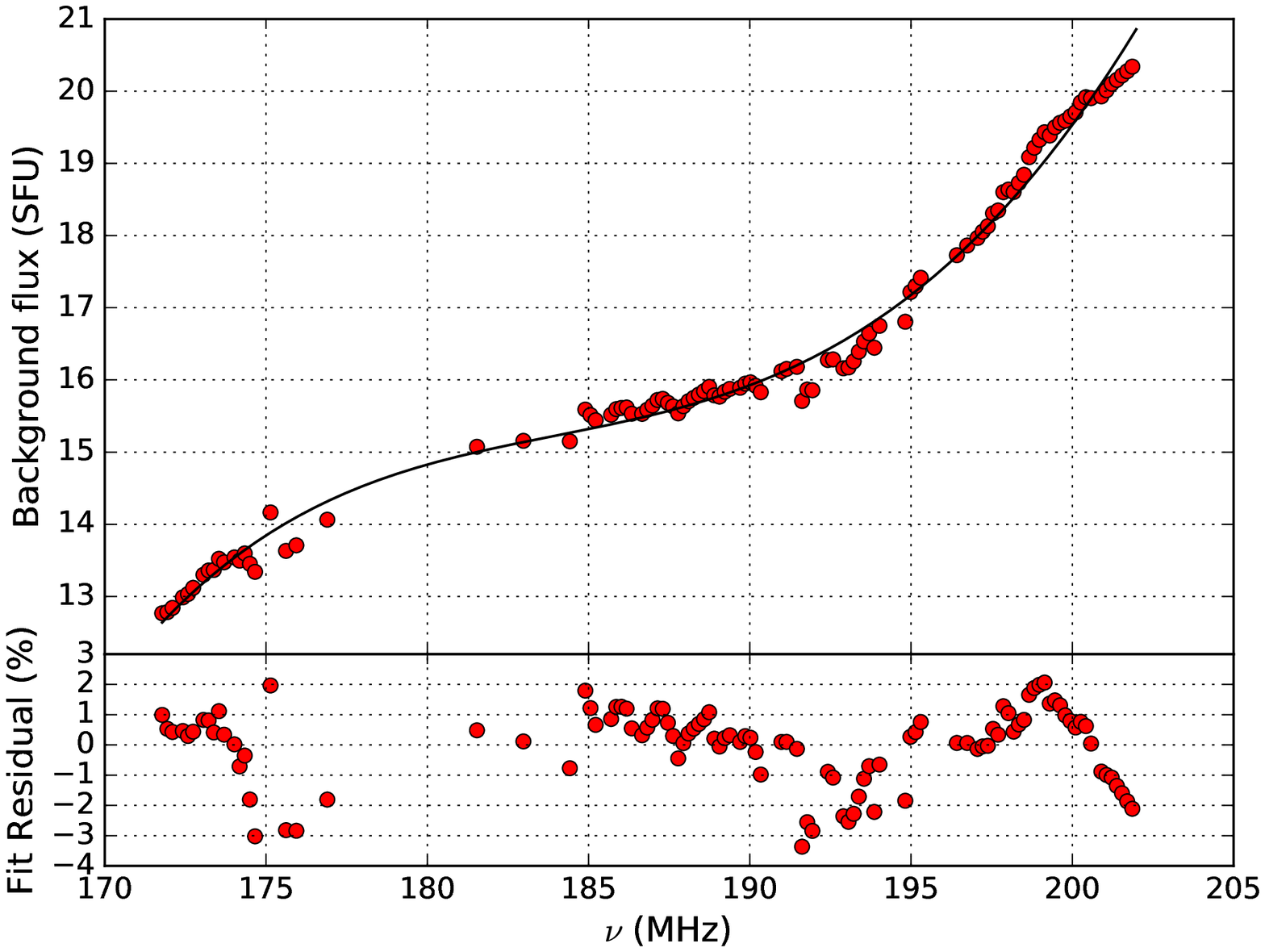}{0.5\textwidth}{(c)}
          \fig{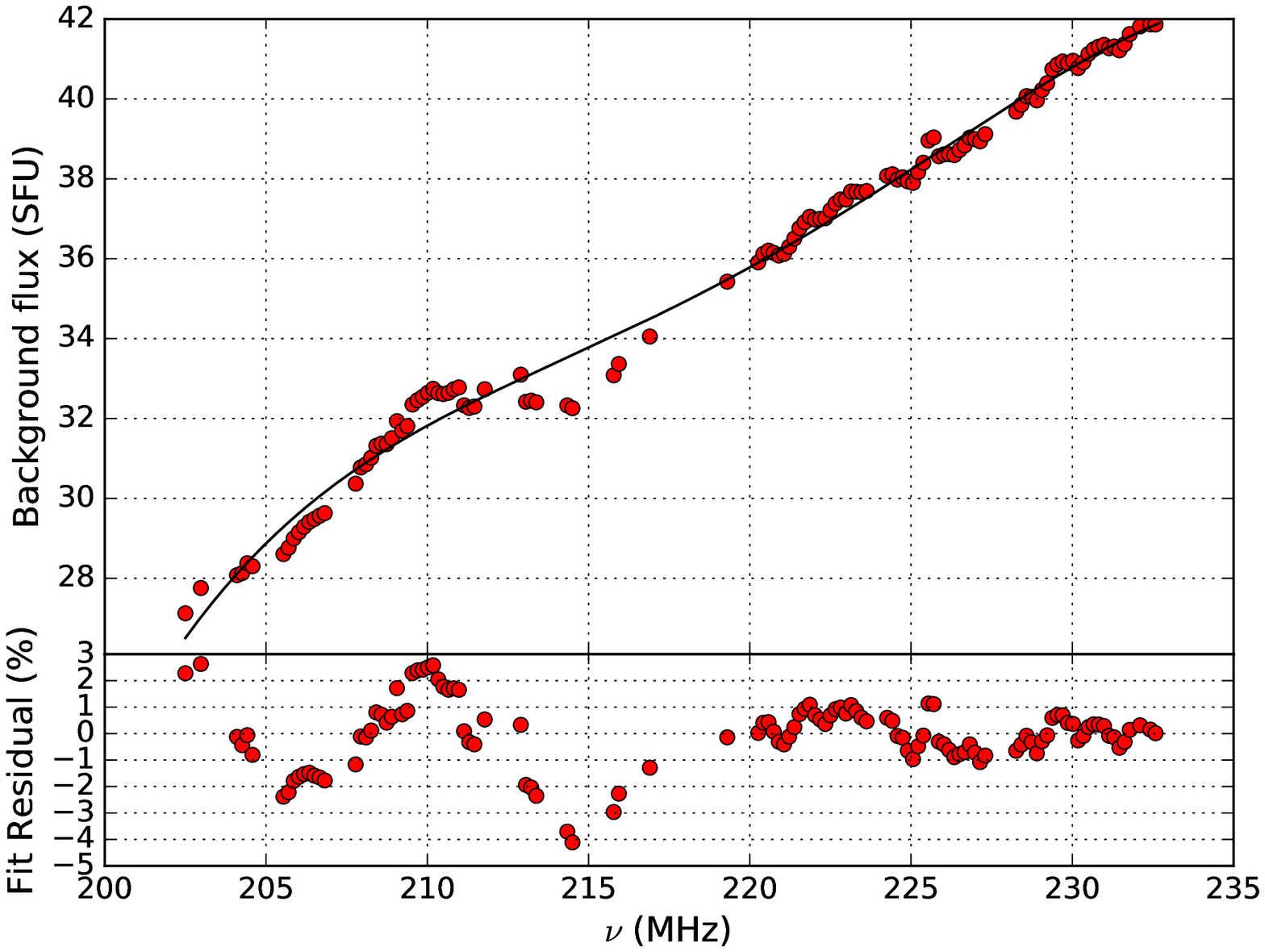}{0.5\textwidth}{(d)}
          }
\caption{Degree-4 polynomial fits to the spectral trend in the background continuum and the residuals to the fits. The 4 panels corresponds to 4 different data sets with frequency ranges: $\text{(a) }110.34-140.54 \text{ MHz}$, $\text{(b) }141.06-171.26 \text{ MHz}$, $\text{(c) }171.78-201.98 \text{ MHz}$, and $\text{(d) }202.5-232.7 \text{ MHz}$. The top sub-panel shows the polynomial fit and the bottom sub-panel shows the departure of the best fit from the data in percentage units. \label{fig:f3}}
\end{figure*}

As the day of our observations was characterized by medium levels of solar activity, it seems reasonable to expect that the thermal quiet Sun emission forms the dominant component of the background continuum emission in our data. We find the temporal variation of the background flux density to be negligible over the duration of individual observing scans of four minutes each. This allows us to then ignore the time dependence of the background flux density and treat it as as a function of frequency alone. As the flux densities of the weakest radio bursts detected in our data sets are only a few percent of the background flux, an accurate and robust means of determining and subtracting out the spectral variation of the background component is required. \\

In this work, the Gaussian Mixtures Model (GMM) routine provided by Scikit-Learn \citep{Pedregosa2011} is applied for estimation of the background flux density $(S_{\odot,B})$ as a function of frequency. As the background is expected to vary smoothly, the DS is divided into contiguous groups of 4 fine spectral channels each. The data in each of these groups is then decomposed as a sum of Gaussians using the GMM routine. As there exists no unique way of representing a given function as a sum of Gaussians, the Bayesian Information Criterion \citep{Burnham2002} has been employed to determine the optimum number of Gaussians required to fit the data. Since the thermal quiet Sun component forms the baseline emission level on top of which non-thermal radio emission is detected, it is reasonable to assume that the Gaussian corresponding to this background continuum must be the one with the lowest mean and the highest weight. For every group of fine spectral channels, the value of the mean of this Gaussian is noted as the background flux density ($S_{GMM} (\nu)$) at the respective frequency and is shown by the red circles in top sub-panels in Fig. \ref{fig:f3}. Presence of strong, frequent radio bursts that outshine the background component in a DS degrade the ability of GMM to determine a value of the background flux at each observing frequency. Our observations were taken on a day with moderate solar activity, allowing for the use of GMM to determine the background flux density at several frequencies in most data sets.\\ 

A degree-4 polynomial is then used to fit the large-scale smooth spectral trend in the background flux density and is subtracted from the DS. The right panel in Fig. \ref{fig:f2} depicts the DS obtained after background removal from the DS depicted in the left panel. The suitability of a degree-4 polynomial fit to the background can be quantified by estimating the residual percentage between $S_{GMM} (\nu)$ and the flux densities ($S_{fit} (\nu)$) predicted from the best fit polynomial at the same frequency. The residual percentage is given by:
\begin{equation}\label{eq1}
\text{Residual \%} (\nu) = \frac{S_{GMM} (\nu) - S_{fit} (\nu)}{S_{GMM} (\nu)} \times 100 \%
\end{equation} 
Figure \ref{fig:f3} depicts the degree-4 polynomial fits to the estimated background fluxes for a few of the DS used in our study. A degree-4 polynomial is adequate to describe the spectral variation observed in the background flux to within a mean absolute error of 3-4\%.

\subsection{Wavelet-Based Feature Detection}\label{wavelet_work}

Continuous wavelet transform (CWT) provides a natural way of obtaining a time-frequency representation of a non-stationary signal through the use of a wavepacket with finite oscillation, i.e, a wavelet. In this work, our signal is the 2D MWA DS containing the features of interest. The efficiency of CWT at reliable detection of  features from the DS depends upon our choice of the 2D mother wavelet and is maximized for a mother wavelet which closely matches the shape of the spectral and temporal profiles of these features.
  
\subsubsection{Choice of mother wavelet}\label{sec:mother_wavelet}
\begin{figure*}
\figurenum{4}
\epsscale{1.16}
\plottwo{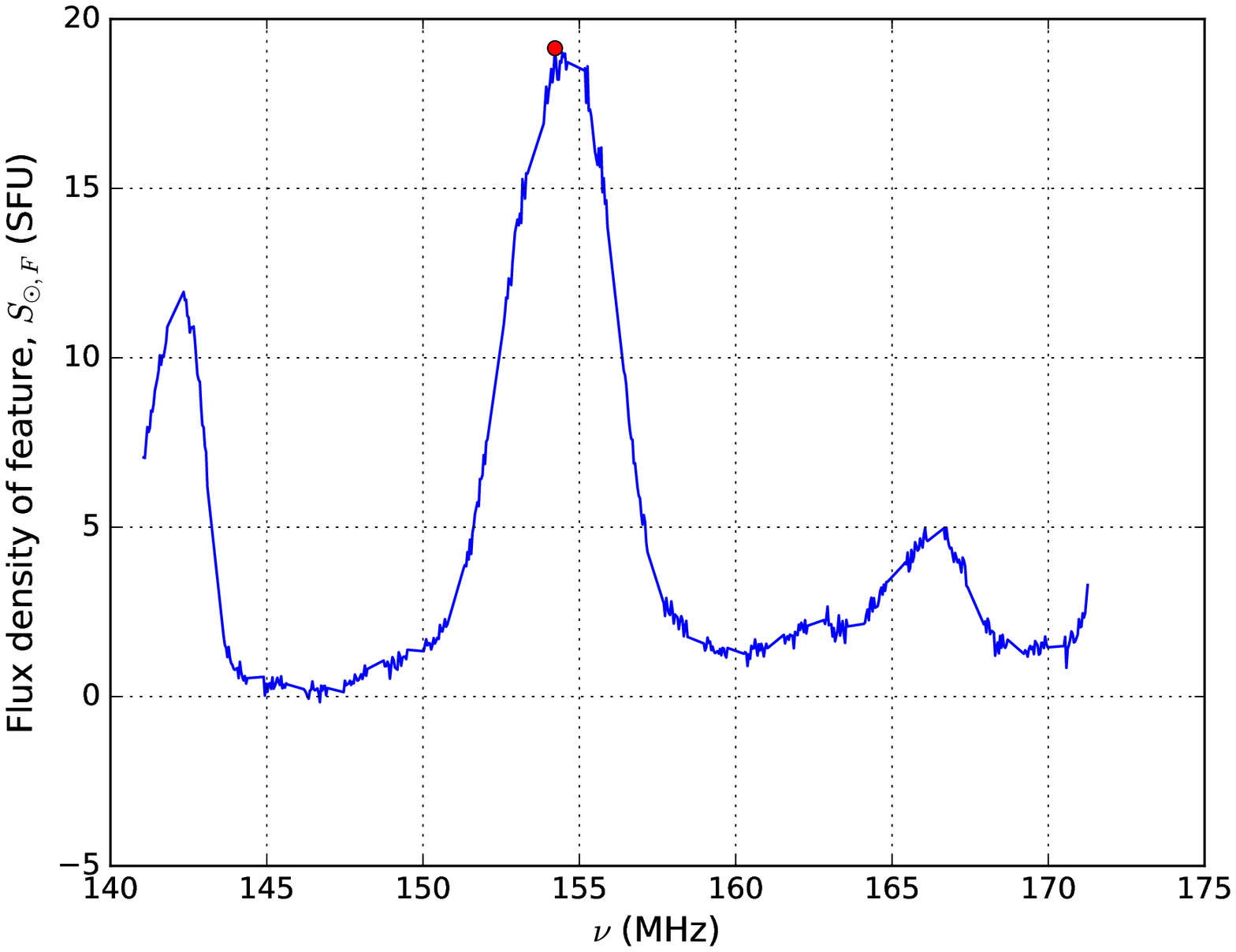}{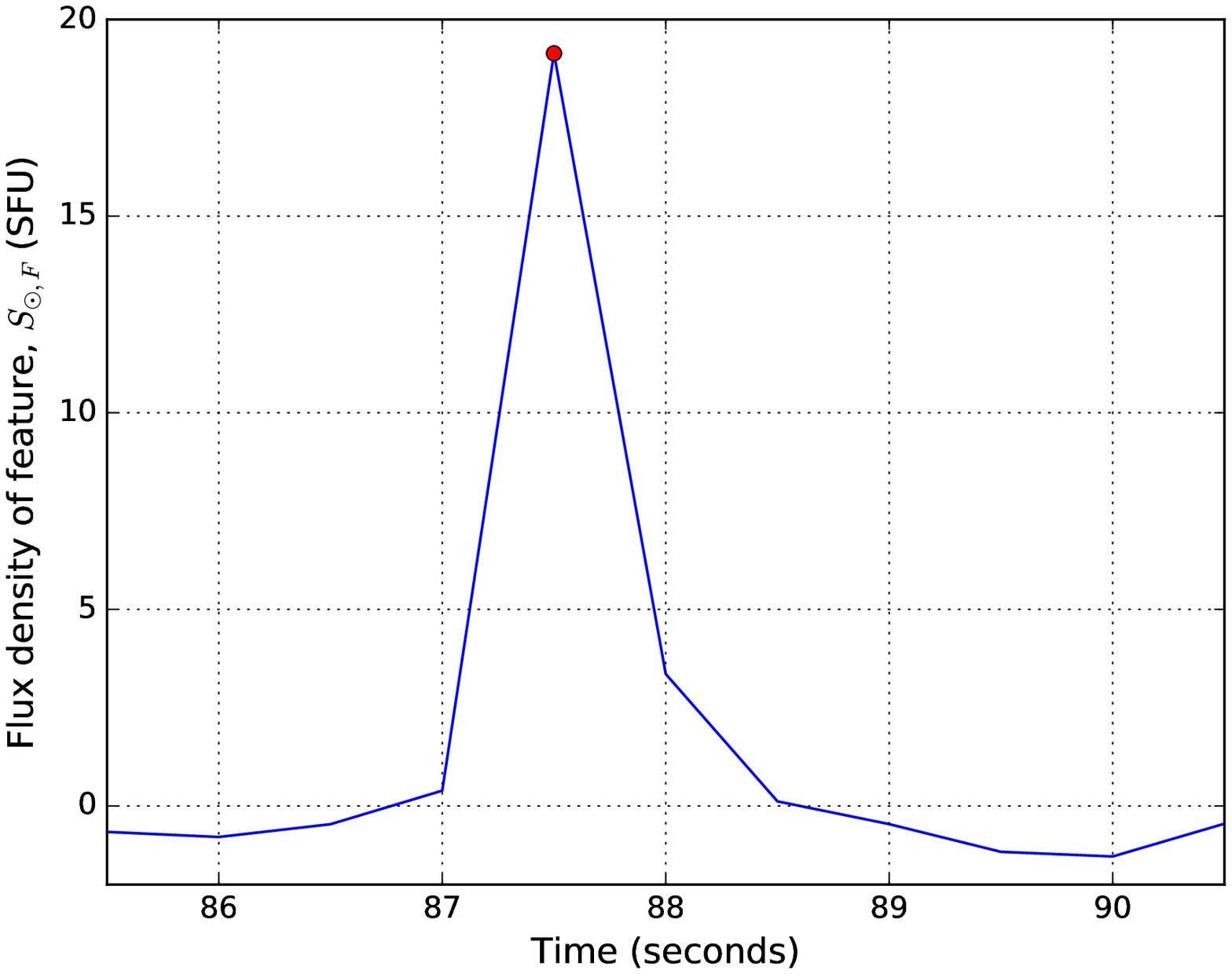}
\caption{Left: Spectral profile of the feature marked by a red circle in the background-subtracted DS depicted in Fig. \ref{fig:f2}. Note that two other weaker features are present at the upper and lower frequency ends in the time slice corresponding to the peak of this feature. Right: Temporal profile of the same feature. The red circles in the left and the right panels of this figure mark the location of the feature peak, as shown in the right panel of Fig. \ref{fig:f2}, along the frequency and time axes respectively.
\label{fig:f4}}
\end{figure*}

From the right panel in Fig. \ref{fig:f2}, it can be seen that while there do exist features that appear isolated in the DS, several features in the DS appear to bunch together. Figure \ref{fig:f4} depicts the spectral and temporal profiles of one seemingly isolated feature indicated by a red circle in the background-subtracted DS depicted in Fig. \ref{fig:f2}. A close look at such isolated features in the DS reveals a characteristic smooth, unimodal nature to their temporal and spectral profiles. Assuming that each atomic feature in a DS possesses unimodal spectral and temporal profiles, any multi-modal spectral or temporal distribution of flux densities observed can be interpreted as a superposition of contributions from constituent unimodal distributions. This allows for a 2D Ricker wavelet to be employed as a suitable mother wavelet for CWT. Measured in pixel units, the features of interest usually have axial ratios of about 10 - 50. To best match features of this nature, we use a variable separable version of a 2D Ricker (also called the Mexican Hat) wavelet  with analytical form :
\begin{equation}\label{eq2}
R(t,\nu)	 = \frac{4}{3\sqrt{\pi}}\left((1-t^2)e^{-\frac{t^2}{2}}\right)\left((1-\nu^2)e^{-\frac{{\nu}^2}{2}}\right)
\end{equation}
as the mother wavelet. From this mother wavelet, scaled wavelets are constructed according to the definition given by \citet{Antoine2004} as follows:
\begin{equation}\label{eq3}
R_{s_{t},s_{\nu},\tau_{t},\tau_{\nu}}(t,\nu)	 = \frac{1}{\sqrt{s_{t}s_{\nu}}}R \left(\frac{t - \tau_{t}}{s_t},\frac{\nu - \tau_{\nu}}{s_\nu}\right)
\end{equation}
The peak of a 2D scaled Ricker wavelet is located at $(t,\nu) = (\tau_{t},\tau_{\nu})$. The scales $s_{\nu}$ and $s_t$ correspond to half the support of its positive lobe along the frequency and time directions respectively. Using the scaled 2D Ricker wavelets, wavelet coefficients of the DS are then computed according to the following definition:
\begin{equation}\label{eq4}
\gamma(s_{t},s_{\nu},\tau_{t},\tau_{\nu}) = \iint\limits_{\nu,t} DS(t,\nu) {R}_{s_{t},s_{\nu},\tau_{t},\tau_{\nu}}(t,\nu) \text{dt d}\nu
\end{equation}
\begin{figure*}
\figurenum{5}
\gridline{\fig{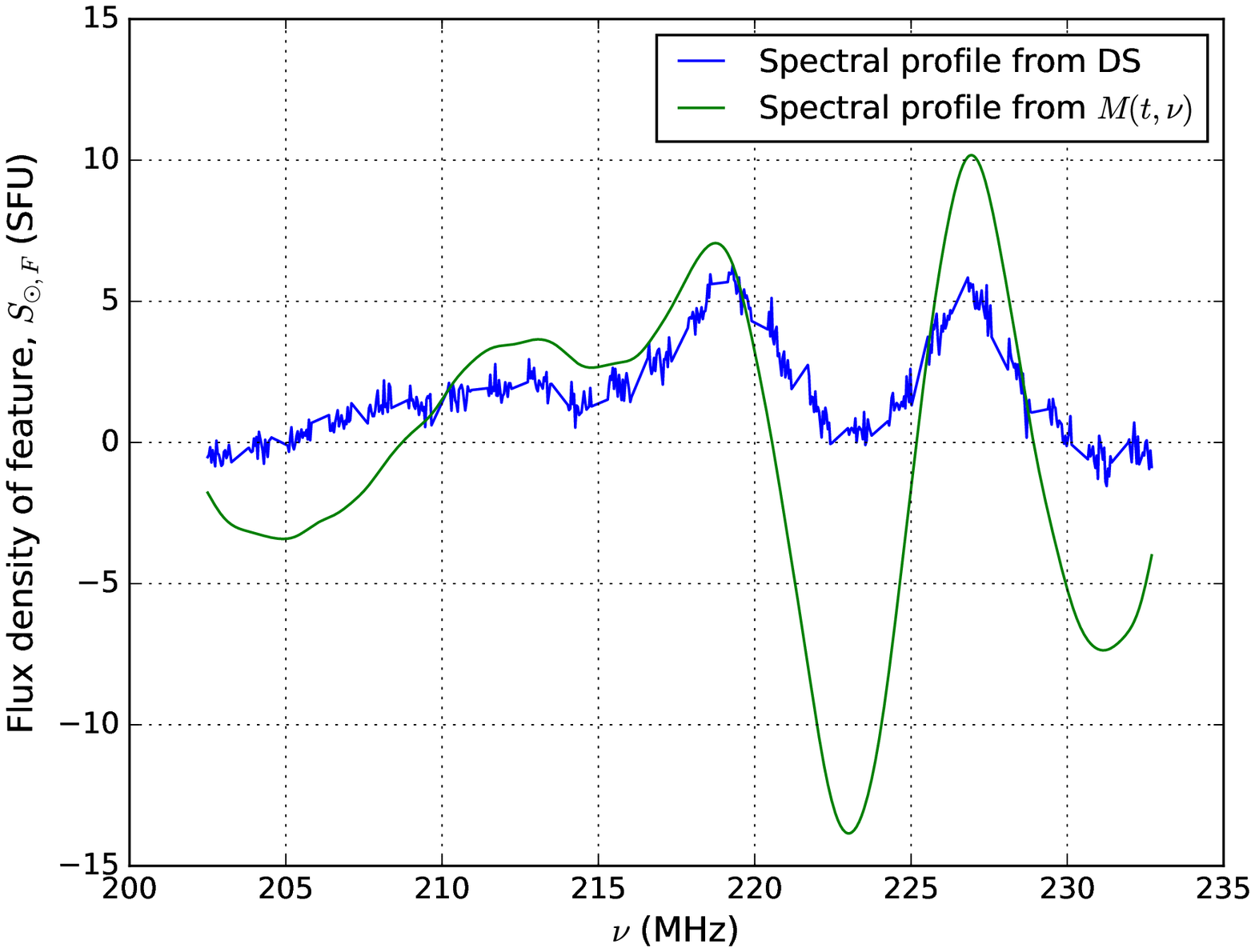}{0.5\textwidth}{(a)}
          \fig{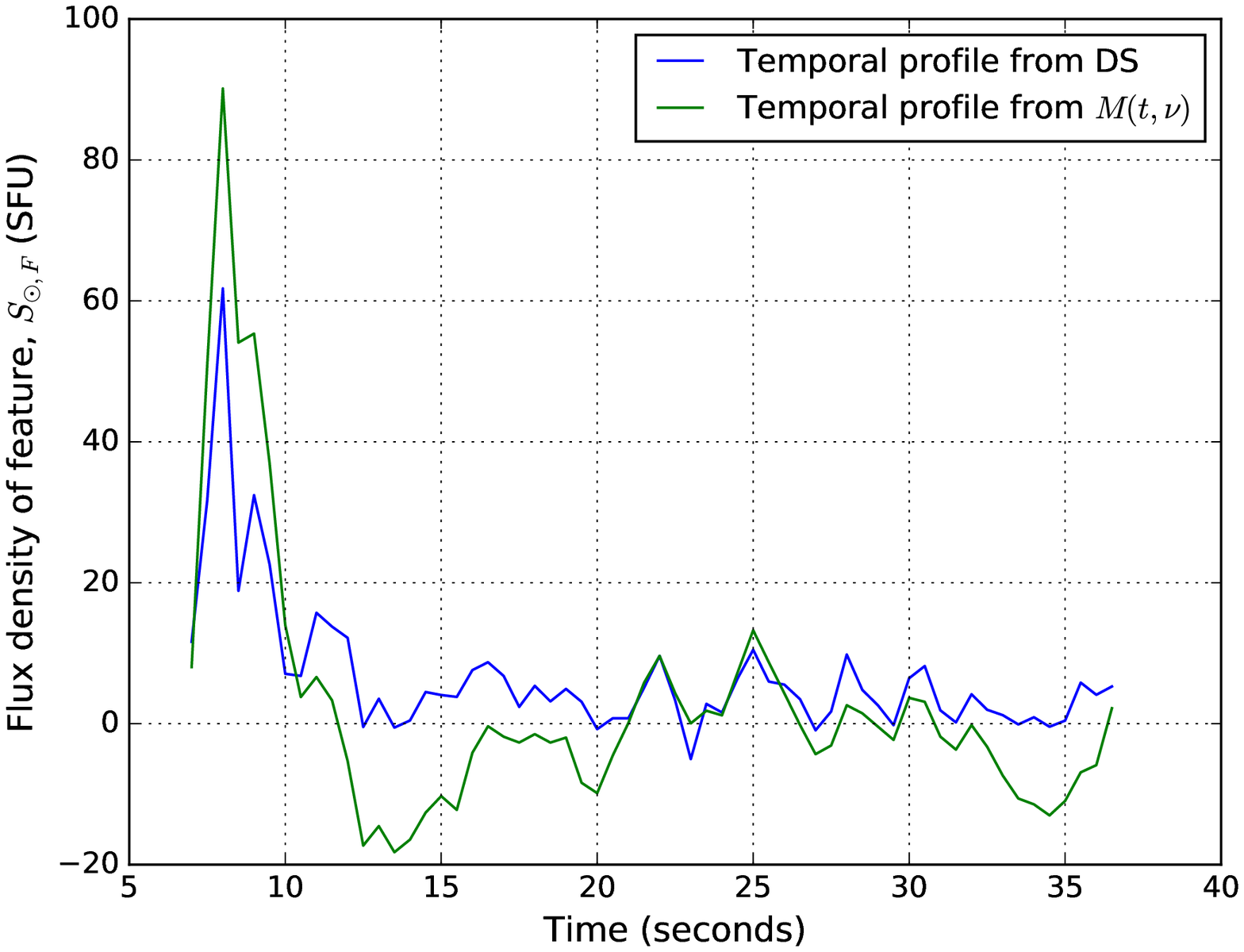}{0.5\textwidth}{(b)}
         }
\caption{(a) A spectral profile taken from both the DS (blue) and $M(t,\nu)$ (green) at the same time. The presence of three local maxima in the $M(t,\nu)$ profile indicates the ability of CWT to successfully detect the three features seen in the DS profile. The locations of the local minima of $M(t,\nu)$ enable us to distinguish individual features from one another despite overlaps in their spectral profiles in the DS. (b) Panel illustrating ability of $M(t,\nu)$ to reproduce temporal widths of features reliably, while  providing the resolution to distinguish between features located close together. The presence of multiple peaks in $M(t,\nu)$, one corresponding to each local maximum in the DS, clearly demonstrates the ability of $M(t,\nu)$ to detect these features.\label{fig:f5}}
\end{figure*}

For ease of notation, let us denote the wavelet coefficients $\gamma(s_{t},s_{\nu},\tau_{t},\tau_{\nu})$ by the symbol $\gamma(s_{t},s_{\nu},t,\nu)$. For a given feature peaked at $(t,\nu)$ in the DS, $\gamma(s_{t},s_{\nu},t,\nu)$ shall be maximized when $s_{\nu}$ and $s_t$ match with the spectral and temporal extents of the feature respectively. Thus, the 2D Ricker wavelet acts as a peak and support detection filter. This then enables us to determine the peak flux densities as well as the temporal and spectral extents of features in the DS. 

\subsubsection{Construction of a composite matrix}\label{sec:composite_matrix}
Owing to the fact that the 2D CWT introduces two additional degrees of freedom through transformation from a 2D DS space to a 4D wavelet-coefficient space, a large number of wavelet coefficients computed for a given DS carry redundant information. The non-orthogonality of a set of scaled Ricker wavelets further preserves this redundancy. This aspect can then be exploited to reconstruct the DS using a basis different from the set of scaled wavelets. \citet{Torrence1998} give an explicit expression for 1D signal reconstruction from the wavelet coefficients using a basis of $\delta-$functions. Extending this formula to the 2D CWT used here, a composite matrix, $A(t,\nu)$, of wavelet coefficients that exactly reconstructs the DS, barring a constant normalization factor, is given by:
\begin{equation}\label{eq5}
A(t,\nu) = \sum\limits_{s_{\nu} > 0} \sum\limits_{s_{t} > 0} \frac{\gamma(s_{t},s_{\nu},t,\nu)}{\sqrt{s_{t}s_{\nu}}} 
\end{equation}
As the wavelet coefficients are nothing but a convolution of the DS with the scaled wavelets, it is expected that  $A(t,\nu)$ should be a smooth reconstruction of the DS. Local maxima in $A(t,\nu)$ then correspond to peaks of features in the actual DS. However, there are two issues with using $A(t,\nu)$ for feature identification. At small scales, our measurements are dominated by noise. As Eq. \ref{eq5} involves a sum over all values of $s_{\nu}$ and $s_t$, it also tries to incorporate the measurement noise in $A(t,\nu)$. Further, bunching of features leading to overlapping of spectral and temporal profiles of adjacent features in the DS can hinder the ability of CWT to resolve two closely spaced features from one another at large values of $s_{\nu}$ and $s_t$. Hence, it is necessary to work with an intermediate range of scales for constructing a composite matrix, $M(t,\nu)$, that captures details of the features of interest while avoiding being influenced by the inherent measurement noise at small scales and bunching of features at large scales. $M(t,\nu)$ is therefore, constructed using the following expression:
\begin{equation}\label{eq6}
M(t,\nu) = \sum\limits_{s_{\nu} = s_{\nu, lower}}^{s_{\nu, upper}} \sum\limits_{s_t = s_{t,lower}}^{s_{t,upper}} \frac{\gamma(s_{t},s_{\nu},t,\nu)}{\sqrt{s_{t}s_{\nu}}}
\end{equation}
In the time domain, the features of interest are already present at the resolution of the data, forcing us to set $s_{t,lower}$ to 0.5 seconds. Careful visual inspection of a large number of DS revealed that there exist few features with bandwidths less than 0.5 MHz, leading us to a choice of 0.5 MHz for $s_{\nu,lower}$. Again, guided by careful visual inspection of several DS, we set $s_{t,upper} = \text{ 3 seconds}$ and $s_{\nu,upper}$ = 5 MHz in order to provide both the ability to detect atomic features present within a bunch of features and the capability to identify relatively long-lived or broadband features. The values chosen for $s_{t,upper}$ and $s_{\nu,upper}$ in fact enable us to reliably reconstruct features with spectral and temporal extents as large as 26.04 MHz and 15 seconds respectively. $M(t,\nu)$ is then computed using the choices of scales mentioned above. Local maxima picked up from $M(t,\nu)$ correspond to locations of the peak flux densities of different features contained in the DS. Figure \ref{fig:f5} illustrates the ability of CWT to distinguish between closely spaced features despite overlaps in their flux density profiles along the frequency and time axes.\\
\begin{figure*}
\figurenum{6}
\epsscale{1.16}
\plottwo{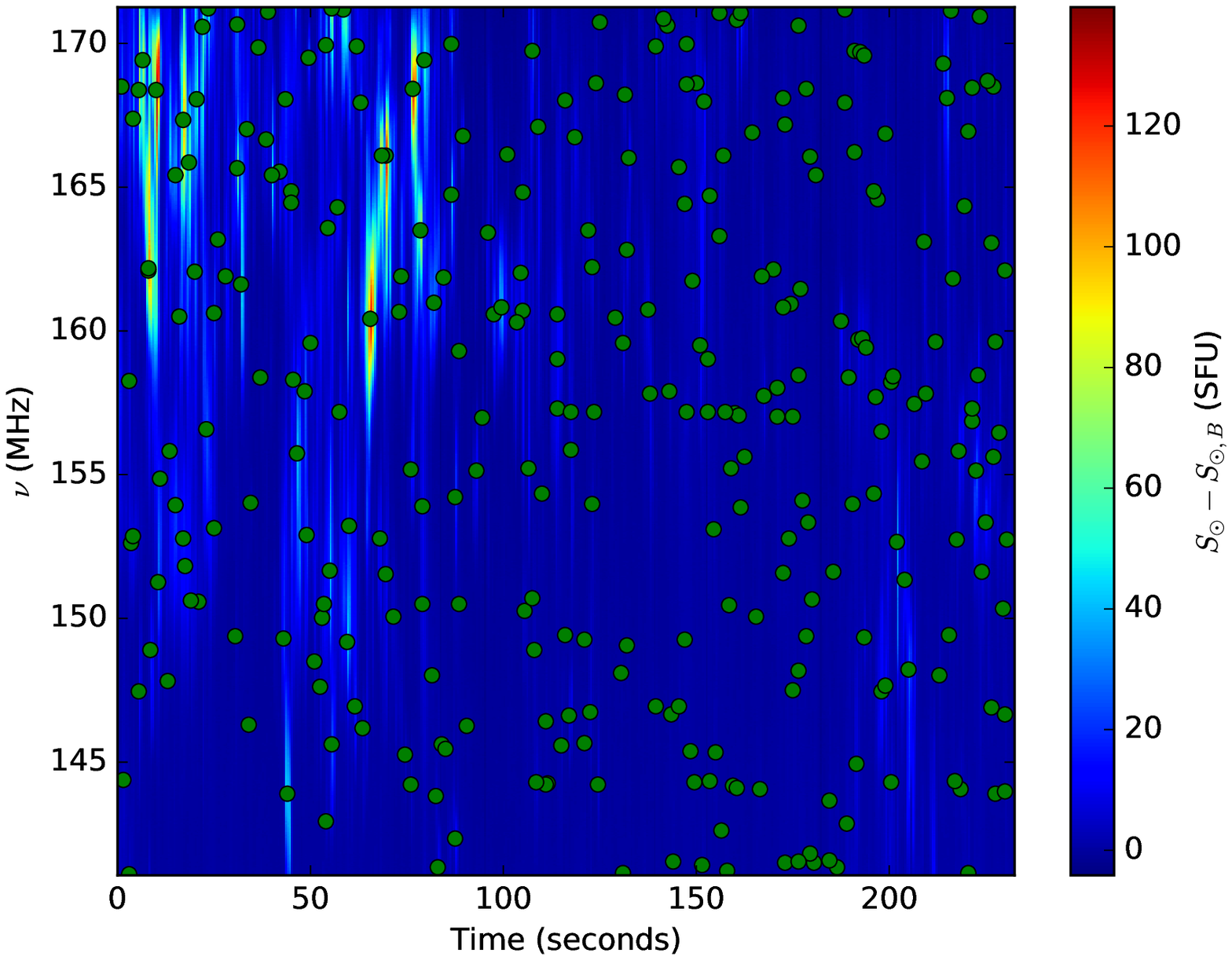}{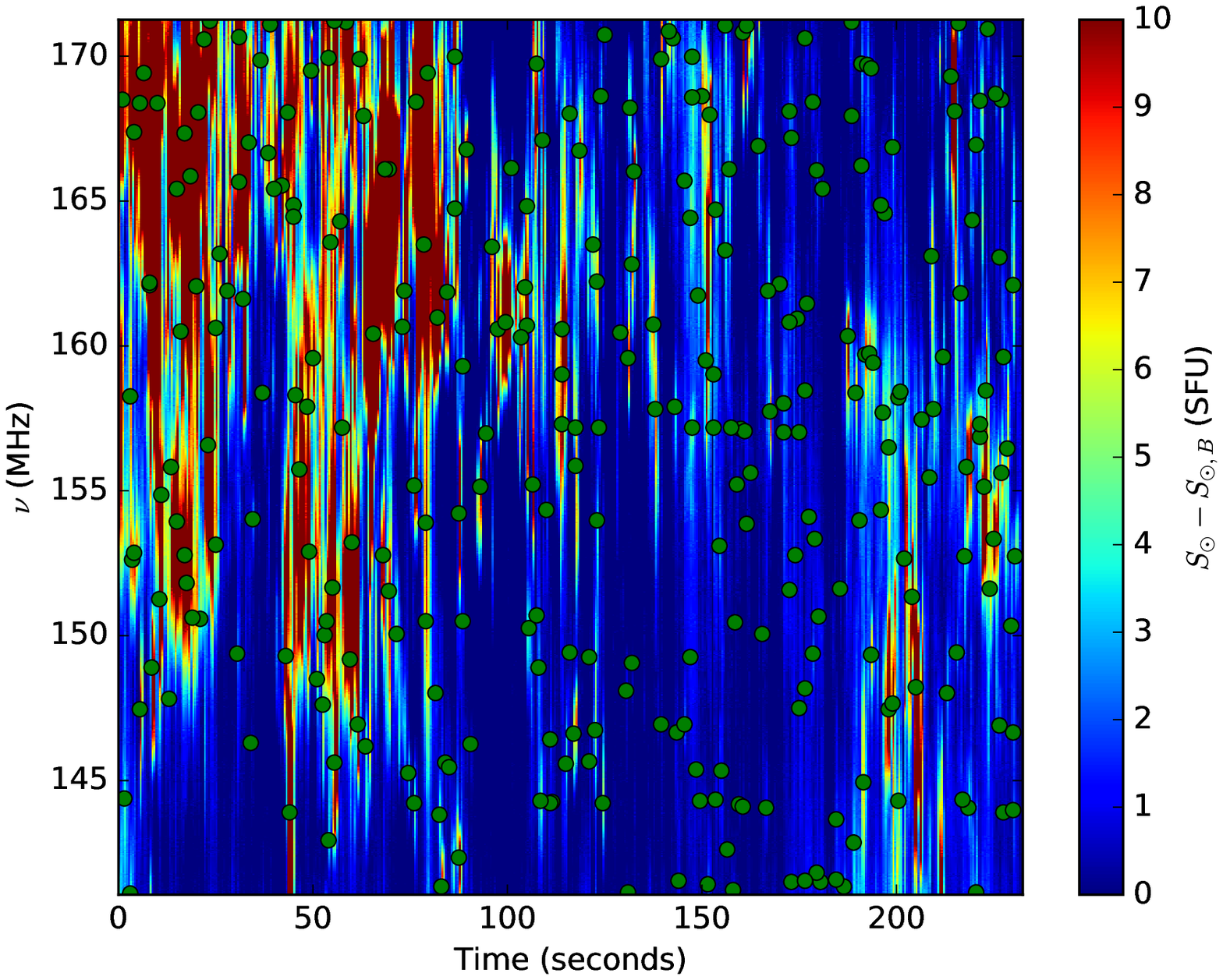}
\caption{Peaks of features detected from the background-subtracted DS depicted in Fig. \ref{fig:f2} are depicted as green circles. The left and the right panels differ only in the color bar range. While the left panel illustrates the ability of the CWT algorithm to pick up bright features, the right panel shows the ability of the CWT code to pick up relatively weaker features as well. \label{fig:f6}}
\end{figure*}

Panel (a) in Fig. \ref{fig:f5} depicts a comparison between a spectral slice taken from both the DS and $M(t,\nu)$ at the same time. The location of peaks of features in the $M(t,\nu)$ spectral profile closely agree with their corresponding peaks in the DS. For a given feature, we find that its spectral extent is matched well by the distance between the two local minima in the $M(t,\nu)$ spectral slice that straddle its peak. We use the distance between these local extrema as the spectral extent of the feature. The lower extremum is then taken to be start frequency ($\nu_{start}$) of the feature. The temporal extent and start time ($t_{start}$) of a feature are similarly estimated. In order to obtain estimates of a quantity similar to the half power width of a feature, we define the spectral and temporal widths of a feature respectively as:
\begin{align*}
\Delta \nu &= 0.5 \times \text{Spectral extent of feature} \\
\Delta t &= 0.5 \times \text{Temporal extent of feature.}
\end{align*}
For the purpose of quantifying any symmetry present in the spectral profile of a feature with peak at frequency $\nu$, we define its spectral symmetry parameter as follows:
\begin{equation}\label{eq7}
\chi_{\nu} = \frac{\nu - \nu_{start}}{2 \Delta \nu}
\end{equation}
The value of this parameter lies in the range from 0 to 1. A spectral symmetry parameter value of 0.5 for a feature represents a perfectly symmetric frequency profile while departures from 0.5 indicate skewness present in the spectral profile. The temporal symmetry parameter ($\chi_t$) of a feature is similarly defined. 

\subsection{Correction of peaks detected}\label{sec:slope_computation}
As seen from panel (a) in Fig. \ref{fig:f5}, peaks of features picked up from $M(t,\nu)$ do not always coincide with their counterparts in the DS. However, since $M(t,\nu)$ peaks lie close to their corresponding DS peaks, this discrepancy is easily corrected by first growing a region around a $M(t,\nu)$ peak and then, identifying the DS peak within this region. The admissibility criterion used to grow a region S starting from a $M(t,\nu)$ peak is that the wavelet coefficient of the neighboring pixel under consideration is within a minimum threshold (T) percentage of the peak wavelet coefficient. The region growing algorithm terminates when no more pixels on the boundary of S satisfy this criterion. Since $M(t,\nu)$ is only an approximation to the actual DS, the temporal and spectral profiles of a feature in the DS are reproduced exactly only within a small neighborhood around its $M(t,\nu)$ peak. Hence, a value of T as high as 95\% has been chosen to ensure that all pixels contained in the region S around the peak of a feature actually belong to this feature. For the features detected from  all DS used in this work, $M(t,\nu)$ peaks show average offsets of  0.16 seconds and 0.57 MHz from their corresponding DS peaks. After peak correction, the peak flux density of a feature is obtained from the flux density at the location of its peak in the background-subtracted DS.
 
\subsection{Elimination of false detections}\label{sec:SNR_falsepeakremoval}
Since $M(t,\nu)$ only approximates the DS, it is possible for it to contain some spurious peaks which do not correspond to real features in the DS. Only a peak in $M(t,\nu)$ having a corresponding peak in the DS is regarded to be a real feature. In order to weed out false peaks, the root mean square flux density ($\sigma$) is estimated across quiet patches in the DS as a function of frequency. A Signal-to-Noise Ratio (SNR) for every peak is then defined as the ratio of the peak flux density to the root mean square background noise at the frequency corresponding to the location of the peak. The spectral and temporal profiles of all peaks detected in $M(t,\nu)$ were visually examined using figures similar to Fig. \ref{fig:f5} to check for a corresponding peak in the DS. We find false detections to constitute about 24\% of the total number of peaks detected in $M(t,\nu)$, all of which have peak flux densities, $S_{\odot,F} < 5\sigma$.  In all, about 26\% of our detections lie below the 5$\sigma$ threshold. In order to eliminate all false positives, we reject all peaks with $S_{\odot,F} < 5\sigma$. \\

Figure \ref{fig:f6} depicts the locations of the peaks of all features detected using this automated wavelet-based approach. In order to estimate the efficiency of this approach at picking up features reliably from the DS, 8 laypersons were presented with plots of different background-subtracted DS similar to Fig. \ref{fig:f6} and requested to estimate the false positive and false negative rates. According to their estimates, the CWT pipeline successfully picks up features from the DS with a zero false positive rate but with a false negative rate of about 4--6\%. A total of 14,177 features were detected from 67 background-subtracted DS used for this work. 

\section{Results}\label{sec:results}
The wavelet-based analysis, yielding a large number of features, allows us to build statistically stable distributions of their properties - their peak flux densities and morphology in the DS. The following sub-sections present the distributions of various quantities of physical interest for these features.

\subsection{Peak flux densities of features}\label{sec:results_fluxes}
\begin{figure}
\figurenum{7}
\epsscale{1.16}
\plotone{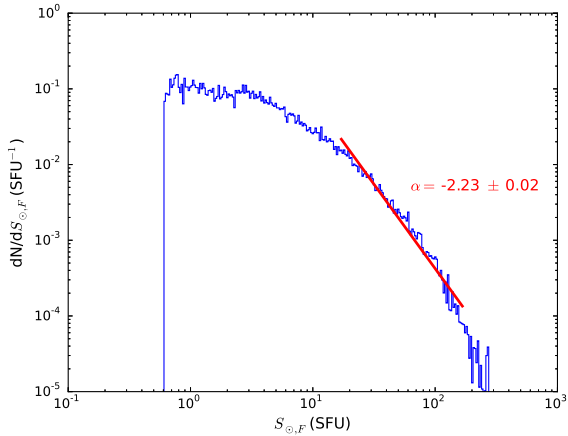}
\caption{Histogram of peak flux densities on a log-log scale. \label{fig:f7}}
\end{figure}
Figure \ref{fig:f7} shows the histogram of peak flux densities ($S_{\odot,F}$) of features. While this histogram extends upto nearly $307 \text{ SFU}$ at its upper end, it touches peak flux densities as low as $0.6 \text{ SFU}$ at its lower end. This makes the detected small-scale features about 1.6 times weaker than the type-I bursts studied by \citet{Ramesh2013} and hence, places them among the weakest reported bursts in literature. A least-squares power law $(\text{dN}/\text{d}S_{\odot,F} \propto {S_{\odot,F}}^{\alpha})$ fit to this histogram yields a power law index $\alpha = -2.23$ over the $12-155$ SFU range. The flux range for this power law fit overlaps with that of the power law fits to the flux density profile done in \citet{Mercier1997}. The upper end of this flux range approaches the lower end of the flux range for the power law fits by \citet{Saint-Hilaire2013}. The value of $\alpha$ obtained here is intermediate between the corresponding values obtained by \citet{Saint-Hilaire2013} ($\alpha \approx -1.66 \text{ to} -1.8$) and \citet{Mercier1997} ($\alpha \approx -2.9 \text{ to} -3.6$). \citet{Iwai2014}, in their observational studies of type-I bursts, also report a power law index of $-2.9 \text{ to} -3.3$. The value of $\alpha$ obtained here is also much lower than the power law index of $-3.5$ predicted for the low energy part of the statistical flare spectrum by \citet{Vlahos1995}.  However, it agrees well with the the power law indices, $\alpha  \approx -2.5$ obtained by \citet{Mugundhan2016} and $\alpha \approx -2.2 \text{ to} -2.7$ obtained by \citet{Ramesh2013} in separate studies of type-I bursts observed using the Gauribidanur Radio Observatory. \\
\begin{figure}
\figurenum{8}
\epsscale{1.16}
\plotone{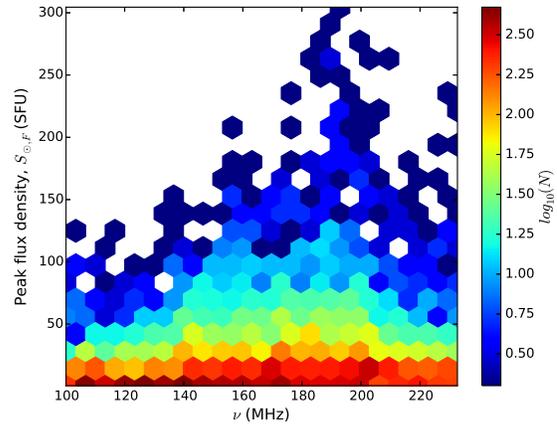}
\caption{Two-dimensional histogram depicting distributions of peak flux densities and peak frequencies. The color axis is in $log_{10}$ units. \label{fig:f8}}
\end{figure}
\begin{figure*}
\figurenum{9}
\epsscale{1.16}
\gridline{\fig{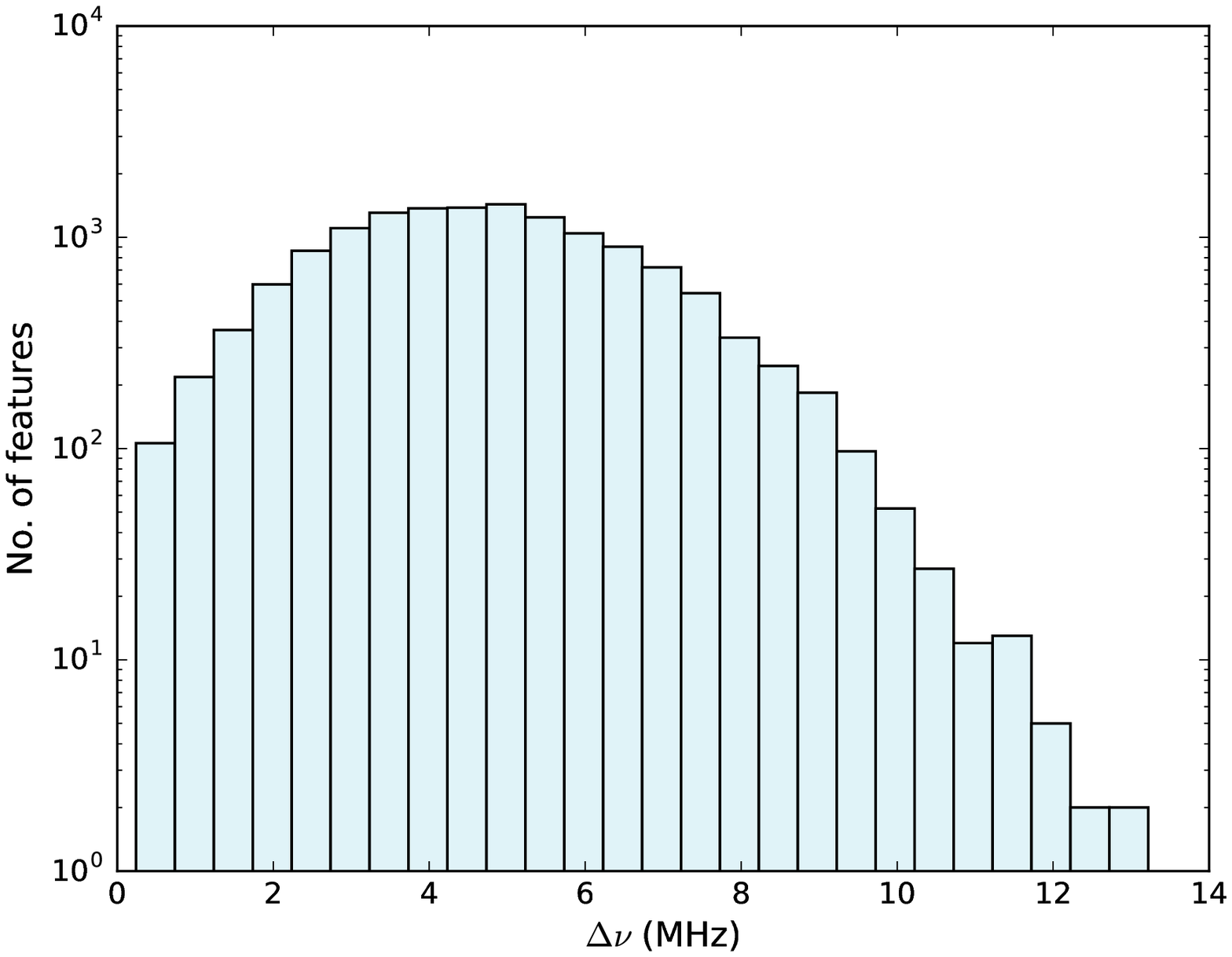}{0.5\textwidth}{(a)}
          \fig{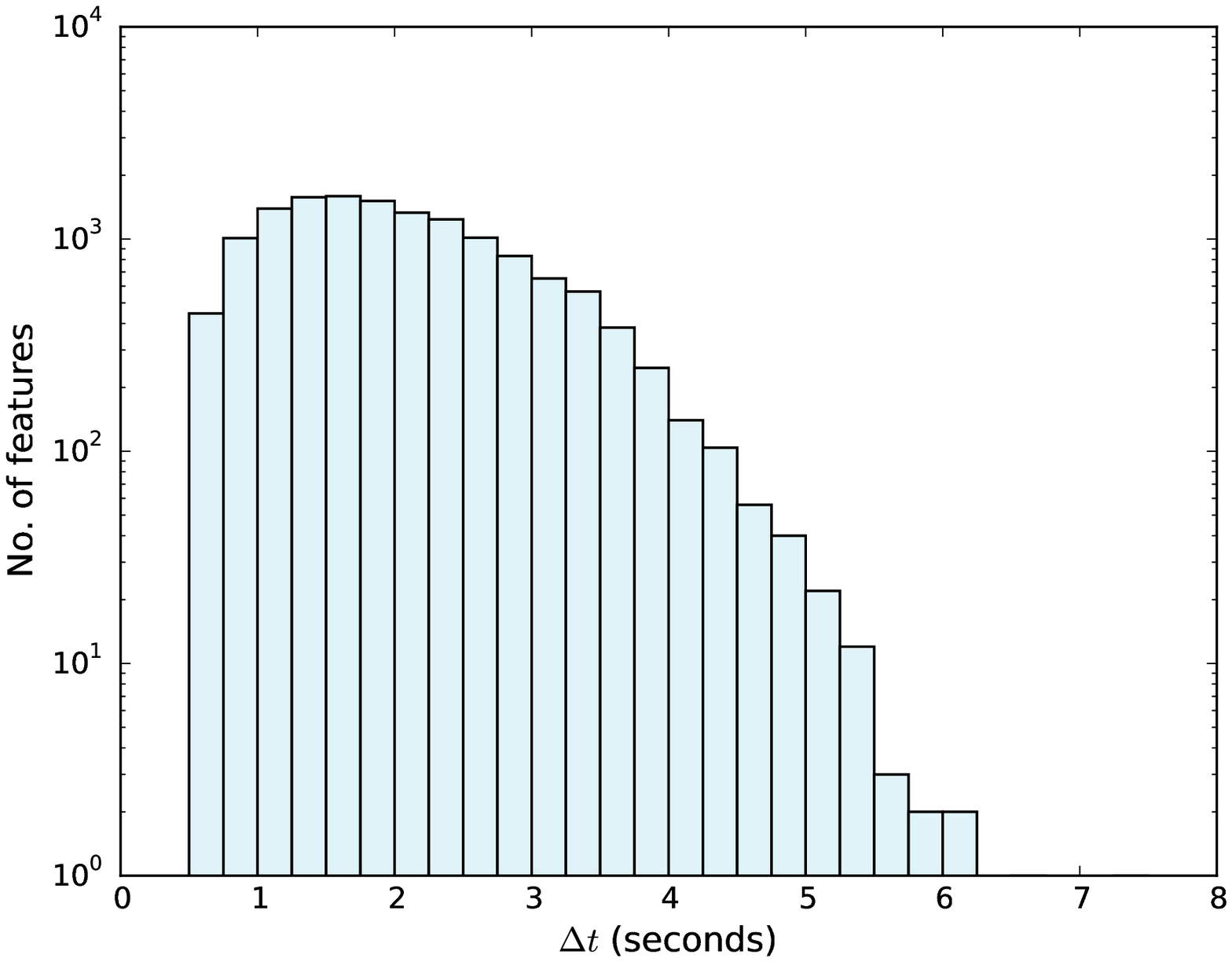}{0.5\textwidth}{(b)}
          }
\gridline{\fig{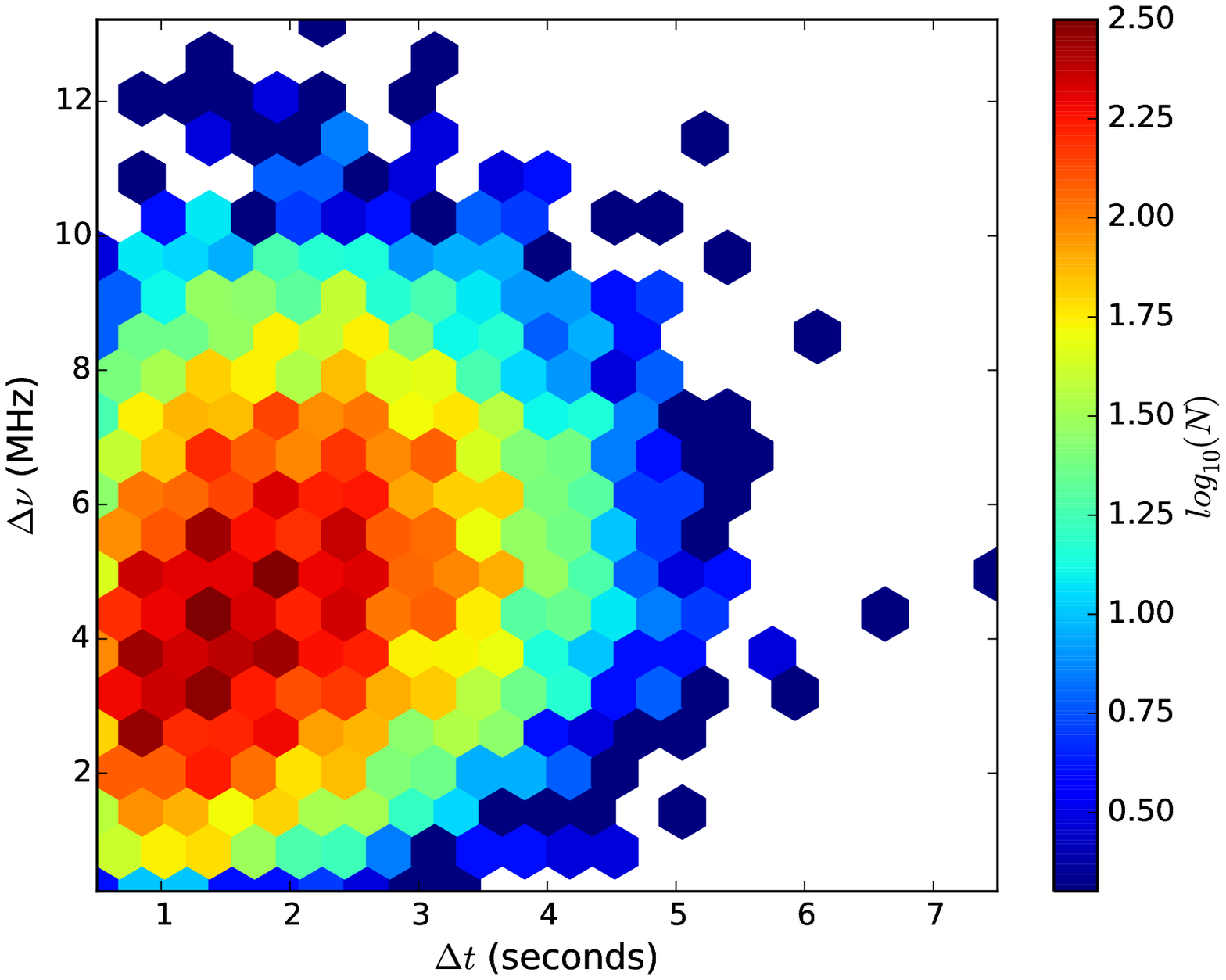}{0.5\textwidth}{(c)}
          }
\caption{(a) Histogram of spectral widths, $\Delta \nu$. (b) Histogram of temporal widths, $\Delta t$. (c) Two-dimensional histogram showing the distributions of $\Delta \nu$ and $\Delta t$. The color axis is in $log_{10}$ units. \label{fig:f9}}
\end{figure*}

We note that the residuals to the power law fit in Fig. \ref{fig:f7} are non-Gaussian, implying the inadequacy of the power law model to fit these data. The uncertainty in the best fit power law slope is, however, only about 1\%, implying that it still provides a reasonable, if sub-optimal, description of the distribution. While a higher order polynomial in log-log space would provide a better fit, we have chosen to use a power law model as it renders itself to an interesting physical interpretation from a coronal heating perspective (Sec. \ref{sec:energy_features}) and provides a point of comparison with earlier literature in the field. We note that the distribution of peak flux densities depicted in Fig. \ref{fig:f7} suffers from  incompleteness at low flux densities and  limited statistics at high flux densities. We have, hence, chosen to fit the power law to an intermediate range of flux densities where the obtained histogram is expected to resemble the true distribution. Though the numerical value of the power law index depends on the exact choice of endpoints chosen for the power law fit, the index is found to be less than -2 irrespective of this choice in the flux density range $\sim 10-160 $ SFU, where we expect the distribution to be complete.\\

Figure \ref{fig:f8} shows a two-dimensional histogram of the distributions of the peak flux densities and the peak frequencies($\nu)$ of the features. While the peak flux densities of a majority of features appear to be independent of $\nu$, a sub-population of them seem to show a frequency-dependent variation in the peak flux density. For this sub-population, the peak flux density appears to increase with $\nu$ from $100 \text{ MHz}$ to $150 \text{ MHz}$, remain nearly constant with $\nu$ between $150 \text{ MHz}$ and $200 \text{ MHz}$, and then decline for $\nu \geq 200 \text{ MHz}$.
\begin{figure*}
\figurenum{10}
\epsscale{1.16}
\plottwo{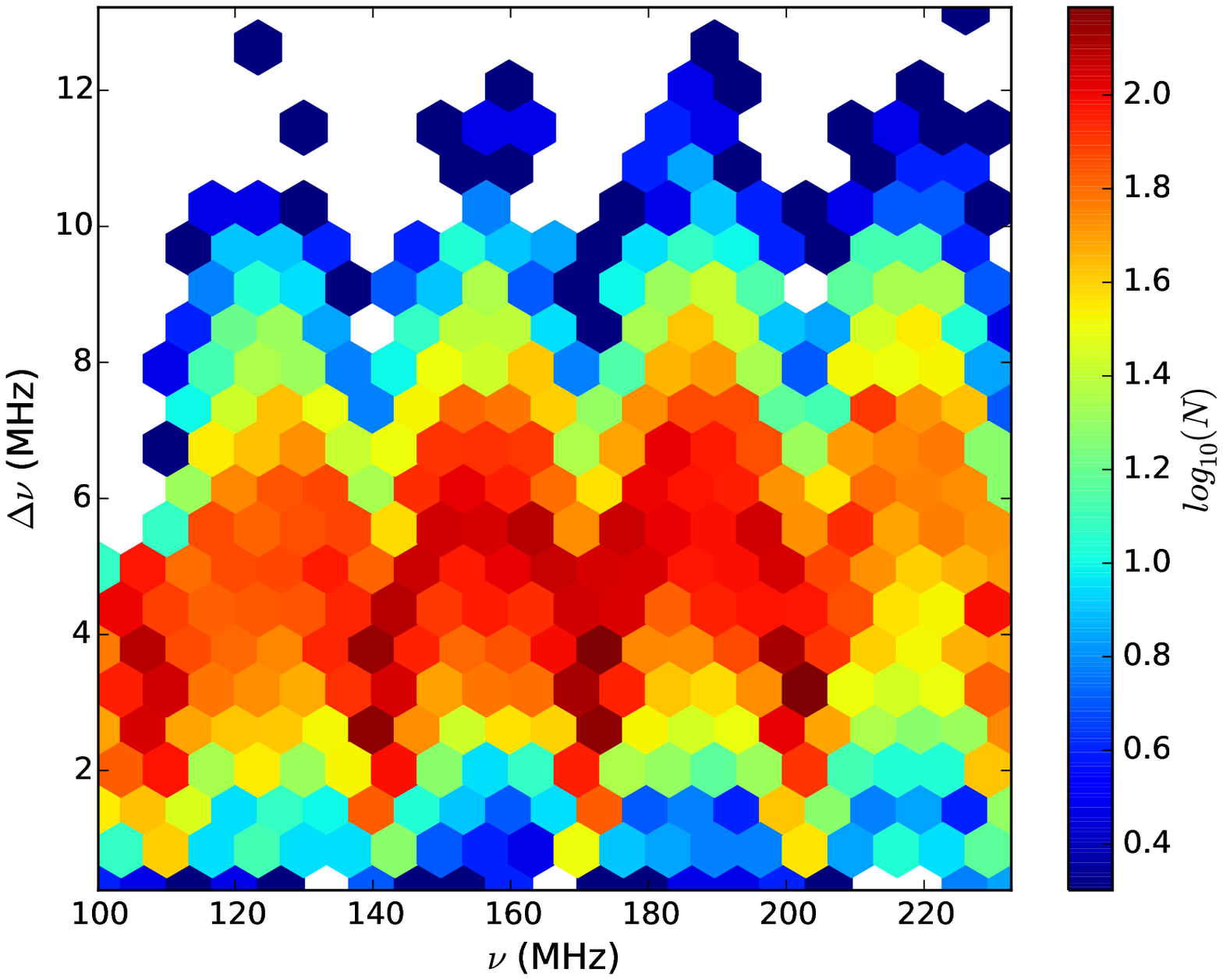}{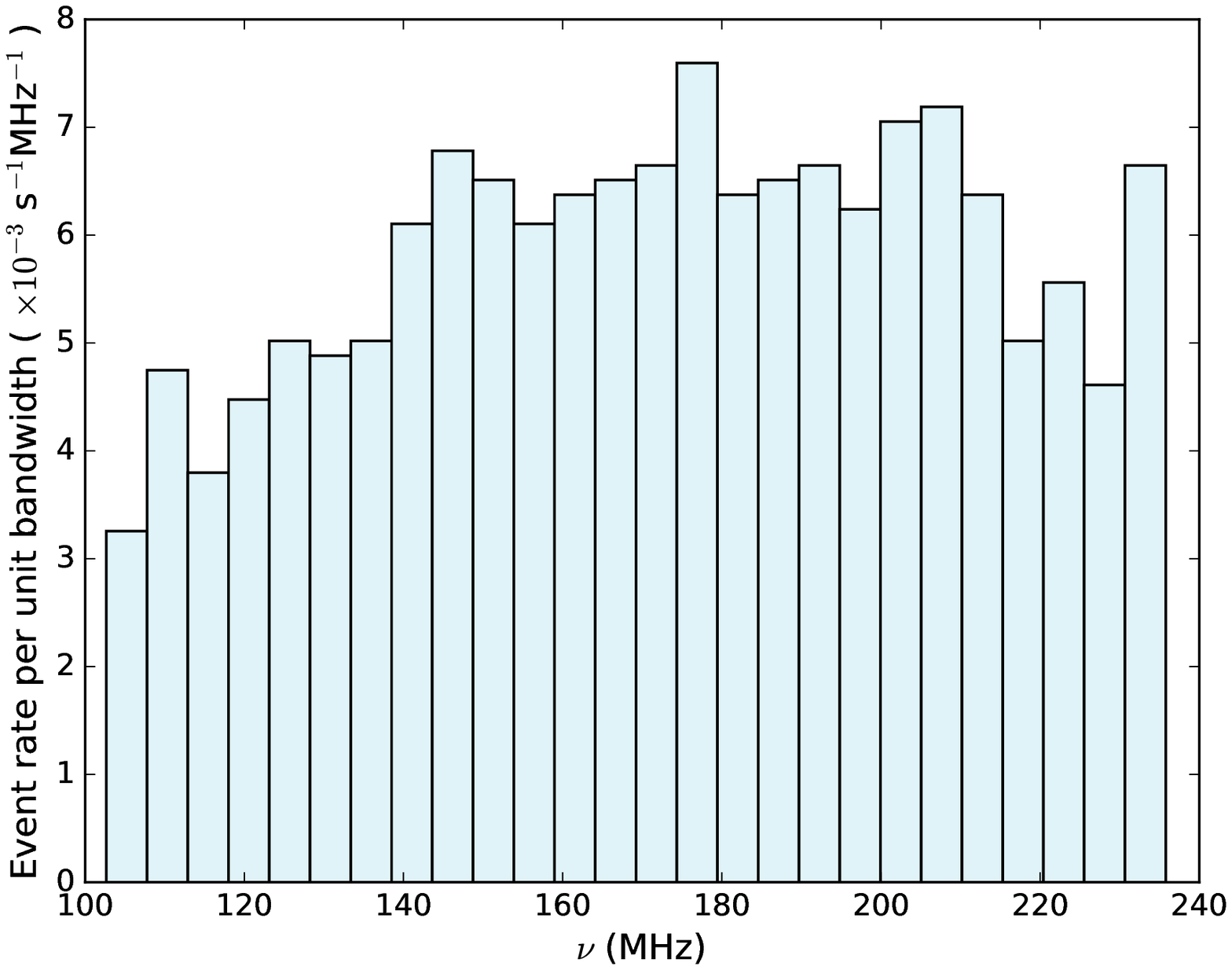}
\caption{Left: Two-dimensional histogram showing the distribution of peak frequencies, $\nu$, and spectral spans, $\Delta \nu$. The color axis is in $log_{10}$ units. Right: Histogram of feature occurrence rate per unit bandwidth.\label{fig:f10}}
\end{figure*}
\begin{figure*}
\figurenum{11}
\epsscale{1.16}
\plottwo{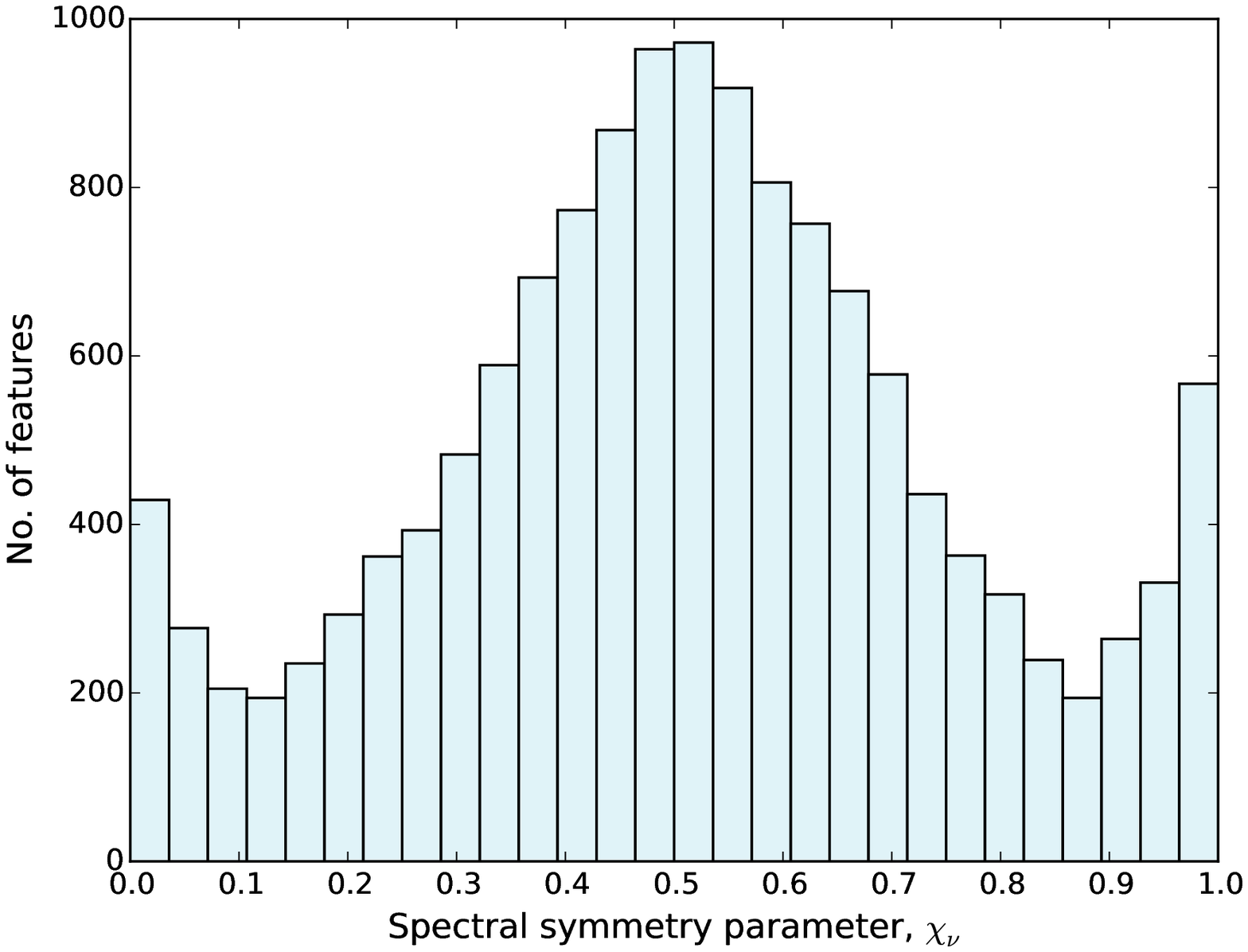}{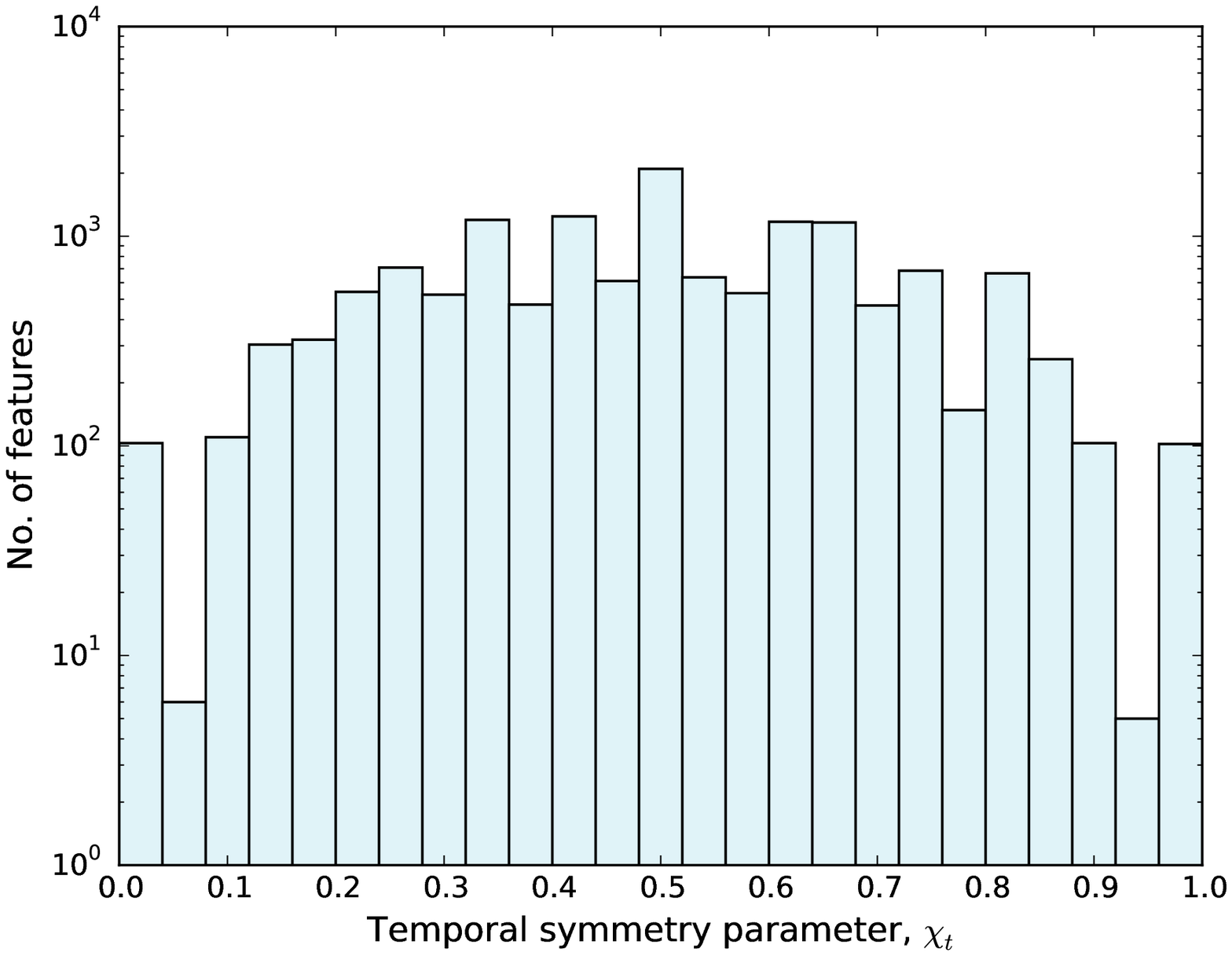}
\caption{Left: Histogram of spectral symmetry parameter. Right: Histogram of temporal symmetry parameter on a semi-log scale.\label{fig:f11}}
\end{figure*}

\subsection{Spectral and temporal widths}\label{sec:results_nu_t}

Figure \ref{fig:f9} depicts histograms of the spectral and temporal widths of features. Both $\Delta \nu$ and $\Delta t$ follow smooth, unimodal distributions. The $\Delta \nu$ distribution peaks at about 4-5 MHz, well above the 40 kHz frequency resolution of our data. On the other hand, the peak in the $\Delta t$ distribution lies at 1-2 s, quite close to the 0.5 s temporal resolution of these data. Fig. \ref{fig:f9}(c) further shows that the distributions of $\Delta \nu$ and $\Delta t$ arrange themselves in a single well-formed cluster peaking at about 4-5 MHz and 1-2 seconds. While the bandwidths of these features are two orders of magnitude smaller than that for type-III bursts, they are comparable to the typical frequency span, $\Delta \nu \lesssim$ 10 MHz \citep{Mercier1997}, reported for type-I bursts.\\
\begin{figure}
\figurenum{12}
\epsscale{1.16}
\plotone{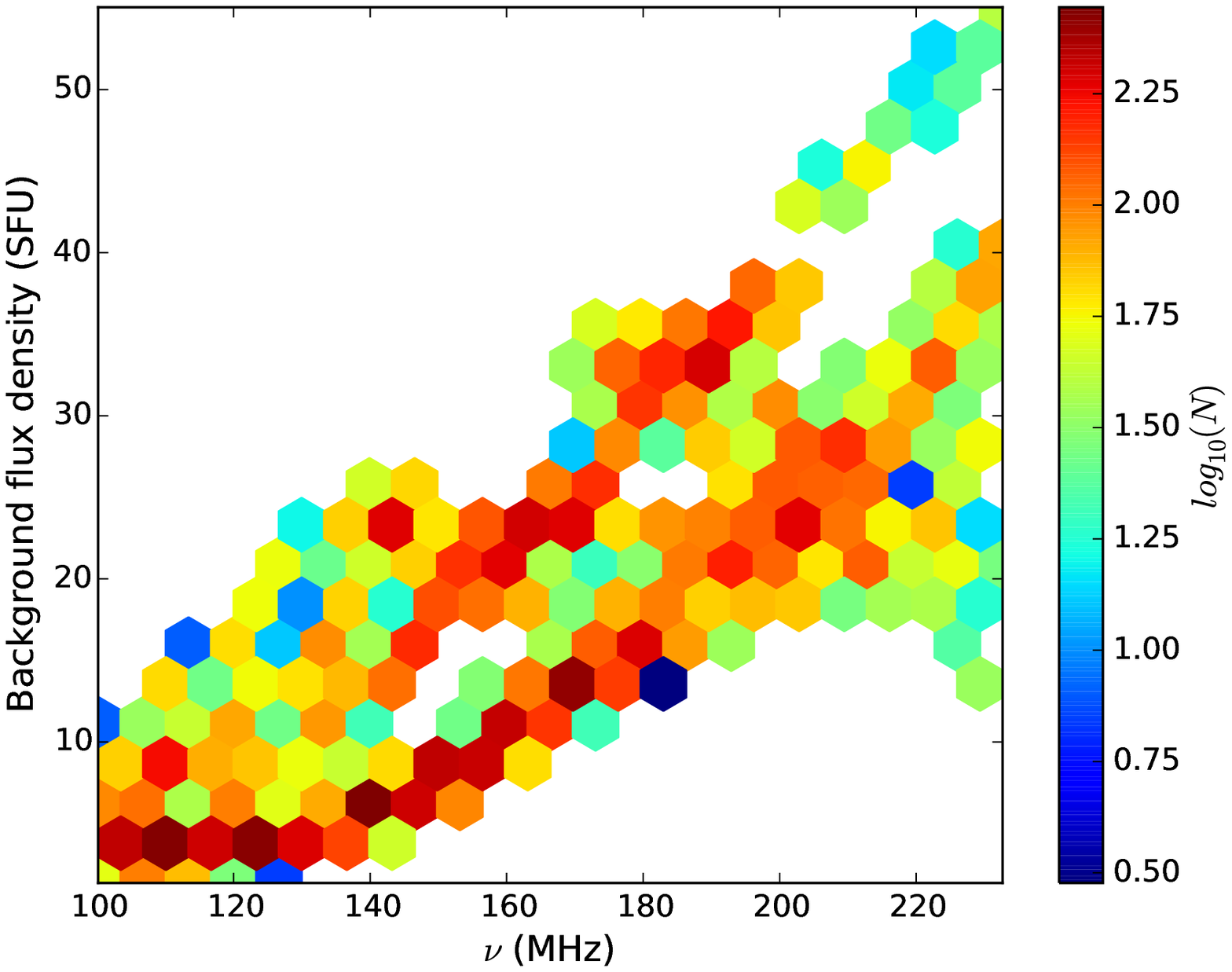}
\caption{Two-dimensional histogram showing distributions of the background flux densities at the locations of peaks in the DS and the corresponding peak frequencies. The color axis is in $log_{10}$ units.\label{fig:f12}}
\end{figure}

The left panel of Fig. \ref{fig:f10} shows a two-dimensional histogram of the distributions of $\Delta \nu$ and the peak frequency ($\nu$). The prominent peak and valley-like structures are artifacts arising from the limited bandwidth of observations. While valleys occur at the edges of the observing bandwidth, peaks occur at its centre. There seems to be a hint of a trend for a small increase in $\Delta \nu$ with increase in $\nu$ ($\approx 0.02 \text{ MHz}$ increase in $\Delta \nu$ per unit increase in $\nu$). The original data set has an equal number of observations at each of the observing bands. The algorithm used to determine the background continuum is designed for periods of medium or low levels of solar activity and hence, worked effectively for most of the data. However, it was not suitable for about 24.3\% of the data which were characterized by high solar activity (typically periods immediately following occurrences of B and C class flares) and hence, discarded from this analysis. This effectively leads to different observing durations for different observing bands and is reflected in the left panel of Fig. \ref{fig:f10}. In order to arrive at the true spectral distribution of features, the feature occurrence rate per unit bandwidth is computed as a function of frequency. As shown in the right panel of Fig. \ref{fig:f10}, the spectral distribution of features appears to remain flat in the frequency range from 140-210 MHz and declines at lower frequencies. Below 140 MHz, the galactic background temperature rises sharply while the intrinsic solar emission becomes weaker \citep{Oberoi2016}. This leads to a drop in the SNR of our detections at these frequencies and consequently, their being under-represented in the spectral distribution. The true spectral distribution of feature occurrences is expected to be flatter than that shown in the right panel of Fig. \ref{fig:f10}.

\subsection{Spectral and temporal profiles}\label{sec:results_spec_temp_profiles}
An interesting finding about the nature of the spectral and temporal profiles of the features of interest is obtained through the histograms of $\chi_{\nu}$ and $\chi_t$ (defined in section \ref{sec:composite_matrix}) shown in Fig. \ref{fig:f11}. While features largely appear to possess symmetric frequency profiles, their temporal profiles display no inherent symmetry. The peaks at the extremes of the $\chi_{\nu}$ and $\chi_t$ histogram range arise due to the presence of features with peaks located close to the edges of the DS. 
 
\subsection{Background flux densities at peak frequencies}\label{sec:results_bgr}
Figure \ref{fig:f12} shows a two-dimensional histogram of the background flux density as a function of frequency. The background continuum emission shows the expected monotonic increase with frequency due to the broadband thermal radiation from the $10^6$ K coronal plasma. RSTN (\url{http://www.sws.bom.gov.au/World_Data_Centre}) solar flux measurements estimate the median flux density on the day of our observations to be 20 SFU at 245 MHz. We also note that over the course of these observations, the spectrally smooth background flux density is observed to vary by a rather large amount. 

\section{Discussion}\label{sec:discussion}

\subsection{Feature energies}\label{sec:energy_features}

\begin{figure}
\figurenum{13}
\epsscale{1.16}
\plotone{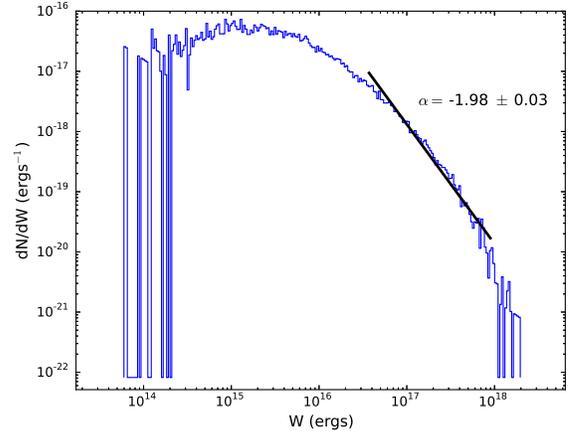}
\caption{Histogram of total energies. \label{fig:f13}}
\end{figure}

Having estimated the peak flux, bandwidth and duration of each feature detected in the DS, estimates of their energy can be obtained if the solid angle into which emission is radiated is known. Assuming isotropic emission for these features, $W = 4 \pi D^2 \Delta \nu \Delta t S_{\odot,F}$ gives the total energy radiated for a feature when observed from a distance $D = $ 1 AU. As this definition of $W$ uses only the peak flux density of a feature without accounting for any reduction in flux density within its shape and assumes isotropic emission, it overestimates the actual energy of a feature. We note that while most earlier works use constant bandwidths and durations to estimate the energy radiated, we use the spectral and temporal spans corresponding to individual features for this purpose. The histogram of total feature energies is shown in Fig. \ref{fig:f14}. The typical energies of these features lie in the range of $10^{15} - 10^{18}\text{ ergs}$. These features are, hence, weaker than both the type-III bursts ($W \approx 10^{18} - 10^{23}$ ergs) investigated by \citet{Saint-Hilaire2013} and type-I bursts ($W \approx 10^{21} \text{ ergs}$) studied by \citet{Mercier1997}.The best fit power law to the tail of the histogram in Fig. \ref{fig:f13} yields a power law index of $-1.98$. \cite{Hudson1991} has shown that for weak flare emissions to play a significant in coronal and chromospheric heating, the power law distribution describing their occurrence must have index $\alpha \leq -2$. Within the uncertainty of the fit, the features studied here meet this criterion, and hence can be expected to play an interesting role in coronal heating. \citet{Subramanian2004} had estimated the ratio of the radiative power output from noise storm continua to the total power input provided to the accelerated non-thermal electrons producing these bursts to be about $10^{-10}-10^{-6}$. Using this efficiency estimate, the typical energies of the non-thermal electrons producing the features of interest lie in the range from $10^{21}-10^{28}$ ergs. This agrees well with the estimate of $10^{23}-10^{26} \text{ ergs}$ obtained by \citet{Subramanian2004} for the energy transferred to the non-thermal electron population that cause noise storm continua. This hints at a possible correlation between the properties of these features with that of type-I bursts. On the basis of observational studies, \citet{Ramesh2013} also report non-thermal electron energies of about  $10^{20} \text{ ergs}$ for radio noise storm bursts.

\subsection{Comparison with type-I bursts}\label{sec:typeI_comparison}
 
Our statistical analysis shows that the features of interest appear to ride on a broadband background continuum. These findings closely agree with observations of type-I bursts present against a continuum emission, giving rise to the speculation that these features might be weak type-I bursts. These results would then support the theory proposed by \citet{Benz1981} and \citet{Spicer1982} that describes type-I bursts as observational signatures of scattered small-scale energy release events in the solar corona. The very electrons accelerated in such small magnetic reconnection events might give rise to the broadband background continuum \citep{Benz1981}. Investigations of type-I bursts in the 160-320 MHz frequency band by \citet{DeGroot1976} suggest an average frequency drift rate of $\text{-10 MHz/s}$ for type-I bursts. The small-scale features detected in the MWA DS appear as vertical streaks with no perceptible frequency drift. However, they might possess small frequency drifts which cannot be measured at the time resolution of the MWA data. \\
\begin{figure}
\figurenum{14}
\epsscale{1.16}
\plotone{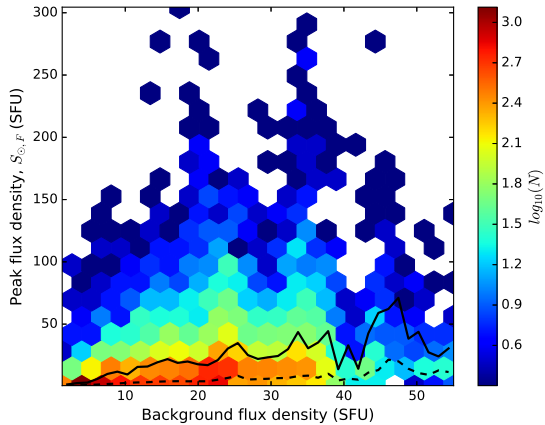}
\caption{Two-dimensional histogram of the peak flux densities of features against the background flux density at their peak frequency. The color axis is in $log_{10}$ units. The dashed and solid black lines represent respectively the first and the third quartile in the distribution of peak flux densities within a background flux density bin. \label{fig:f14}}
\end{figure}

Figure \ref{fig:f14} shows a two-dimensional histogram of their peak flux densities and the background flux density at their peak frequency. The dashed and solid black lines in Fig. \ref{fig:f14} respectively depict trends in the first quartile and the third quartile in the distribution of peak flux densities as a function of the background flux density. The 25th percentile of the distribution of peak fluxes increases from 0.76 SFU at a background flux of 2 SFU to 10.18 SFU at 38 SFU background flux. Similarly the 75th percentile increases from 1.59 SFU to 44.39 SFU over the same range of background flux densities. This demonstrates a tendency for an increase in feature peak flux density with an increase in background flux density. Note that we are limited by statistics at background flux estimates greater than 38 SFU.\\

As shown in Fig. \ref{fig:f12}, the background flux density at any frequency varies by a factor of $\sim$2. Such large variations are seen over time scales as short as 30 minutes and are not likely to reflect changes in thermal emission from the $10^6$ K coronal plasma. The observed increase in the peak flux of the features with increase in background flux density suggests a possibility that this enhanced background continuum could arise due to a superposition of a large number of features which remain unresolved at the time resolution of these data. Observations of such small-scale features with finer time resolution are required to understand them better.  

\subsection{Comparison with type-III bursts}\label{sec:typeIII_comparison}

\citet{Gergely1975} and \cite{Duncan1981} find close similarities between sources of type-I bursts and that of decametric type-III bursts. \citet{Benz1981} claim that electrons accelerated at magnetic reconnection sites, if trapped along closed field lines, produce type-I bursts and their associated continuum. If untrapped, these electrons propagate along open field lines and produce type-III storm bursts. Assuming a type-III-like emission mechanism for the small-scale features observed in the MWA data, we arrive at a one-to-one correspondence between their peak frequencies 
and the electron densities at their heights of production in the solar corona. We assume a 4 $\times$ \citet{Newkirk1961} density profile in the solar corona in order to translate from electron densities \citep{Li2009} to heights ($h$) in the solar corona. Having computed $\nu_{start}$, $\Delta \nu$ and $\Delta t$ for every feature, we can also determine a height band ($\Delta h$) and a propagation speed ($v = \Delta h /\Delta t$)for every feature. Assuming emission at the local plasma frequency, we find that the features of interest mostly possess propagation speeds of about $(0.01 - 0.04)$c and arise in the solar corona from within a narrow band $\Delta h \approx \text{(0.01 - 0.03)} R_{\odot}$ centered at $h \approx \text{(0.20 - 0.50)} R_{\odot}$ above the photosphere. The typical electron speeds drop steadily with frequency, decreasing from $(0.02 - 0.07)$c at 120 MHz to $(0.01 - 0.03)$c at $\text{220 MHz}$. These values, are however, much lower than the speed of 0.33c reported for type-III bursts in the lower corona. As we are unable to discern any spectral drift in these features from the data, the speeds determined here are lower limits to their true speeds.\\

The growth and decay time scales of type-III bursts provide interesting diagnostics for the physical processes involved in their production \citep{Reid2014}. For the small-scale features observed in the MWA DS, the mean and median values of the growth and decay time scales $(\sim 1.5 \text{ s})$ are not found to be significantly different, though we note that they are likely temporally under-sampled by the time resolution of these data.  

\section{Summary, Conclusion and Future Work}\label{sec:conclusion}

We have carried out the first detailed statistical characterization of the small-scale features observed in the MWA solar DS. Owing to their large event rates, it is very hard or impractical to manually attempt to analyze their properties. A robust, automated technique is, hence, necessarily required for our purpose. We have developed a suitable wavelet-based approach to identify, extract and characterize these features. Individual features in the DS possess unimodal spectral and temporal profiles, and a 2D Ricker wavelet is very effective in locating and characterizing them. A total of 14,177 features have been picked up from all DS used in this work. Though our current implementation is adapted for the MWA data, it is quite general and can be applied to DS from other telescopes as well. \\

The CWT algorithm enables us to reliably detect and characterize features with peak flux densities as low as $0.6 \text{ SFU}$, which form the weakest solar radio bursts reported to date in literature. The distribution of their peak flux densities is well-fitted by a power law with index -2.23 over the $12-155$ SFU flux range. We estimate the total radiated energy of these features to be in the range of $10^{15}$ - $10^{18}$ ergs. Hence, they are much weaker than the widely studied solar type-I and type-III bursts. Their energy distribution is fitted well by a power law with index $-1.98$. Within the uncertainty of this fit, this suggests that they could contribute in an energetically significant manner to coronal heating. \\

We find these features to be quite short-lived and narrow-band with typical durations of 1-2 seconds and bandwidths of 4-5 MHz respectively. Interestingly, while their temporal profiles display no structural symmetry, their spectral profiles are largely symmetric about the peak frequency. The distribution of their occurrence rate remains nearly flat in the 140 - 210 MHz frequency range. Quite analogous to type-I bursts, they are also found to reside on an enhanced background continuum. We speculate that these features might correspond to weak type-I bursts. Since type-I bursts and decametric type-III bursts show close associations \citep{Gergely1975,Duncan1981}, it is possible that some of these features could be weak type-III bursts as well.\\

Sensitive high-time resolution observations aimed at searching for a frequency drift and a harmonic counterpart for these features would hopefully provide us with crucial information for understanding them better. Imaging studies to determine their distribution on the solar surface and investigate any correlations with other solar features will further help explore their contributions to coronal heating. We also hope that this detailed and statistically robust characterization of non-thermal emission features will engender interest in the theory and simulation community to understand them better.

\acknowledgments
This scientific work makes use of the Murchison Radio-astronomy Observatory, operated by CSIRO. We acknowledge the Wajarri Yamatji people as the traditional owners of the Observatory site. Support for the operation of the MWA is provided by the Australian Government (NCRIS), under a contract to Curtin University administered by Astronomy Australia Limited. We acknowledge the Pawsey Supercomputing Centre which is supported by the Western Australian and Australian Governments. AS would like to thank Atul Mohan (graduate student at the National Centre for Radio Astrophysics - Tata Institute of Fundamental Research (NCRA-TIFR), Pune, India) for thought-provoking discussions and providing timely, constructive criticisms. AS would also like to thank NCRA-TIFR for the computation and infrastructure support, and the Kishore Vaigyanik Protsahan Yojana scheme under the Department of Science and Technology, Government of India for the financial support provided during the period of this work. 

\facilities{MWA}
\software{Python (\url{http://www.python.org}), NumPy \citep{vanderWalt2011}, Scipy \citep{vanderWalt2011}, Matplotlib \citep{Hunter2007}, Scikit-Learn \citep{Pedregosa2011} and Scikit-Image \citep{vanderWalt2014}. }

\end{document}